\documentclass[showpacs,amsmath,amssymb,prd,superscriptaddress,preprint]{revtex4-1}
\usepackage{amssymb}
\usepackage{graphics}
\usepackage{epstopdf}
\usepackage{epsfig}
\usepackage[thickqspace,squaren]{SIunits} 
  \usepackage{amsmath}
\usepackage{color}
 \definecolor{blue}{rgb}{0,0,1} 
\definecolor{redi}{rgb}{0.7876,0.278,0.0078}
\definecolor{greeni}{rgb}{0,0.6,0}
\usepackage{bm}
\usepackage{tikz}\usepackage{graphicx,nicefrac}
\begin{document}

\title{Dynamical ion transfer between coupled Coulomb crystals in a double well potential}
\author{Andrea Klumpp}
\affiliation{Zentrum f\"{u}r Optische Quantentechnologien, Universit\"{a}t Hamburg, Luruper Chaussee 149, 22761 Hamburg, Germany}
\author{Alexandra Zampetaki}
\affiliation{Zentrum f\"{u}r Optische Quantentechnologien, Universit\"{a}t Hamburg, Luruper Chaussee 149, 22761 Hamburg, Germany}
\author{P. Schmelcher}
\affiliation{Zentrum f\"{u}r Optische Quantentechnologien, Universit\"{a}t Hamburg, Luruper Chaussee 149, 22761 Hamburg, Germany}
\affiliation{The Hamburg Centre for Ultrafast Imaging, Luruper Chaussee 149, 22761 Hamburg, Germany}

\begin{abstract}

 We investigate the non-equilibrium dynamics of coupled Coulomb crystals of different sizes trapped in a double well potential.
The dynamics is  induced by an instantaneous quench of the potential barrier separating the two crystals.
Due to the intra- and inter-crystal Coulomb interactions and the asymmetric population of the potential
wells we observe a complex reordering  of ions within the two crystals as well as ion transfer processes from one well to the other.
The study and analysis of the latter processes constitutes the main focus of this work. In particular we examine the dependence of the observed ion transfers on the quench amplitude performing
an analysis for different crystalline configurations ranging from  one-dimensional ion chains
via two-dimensional zig-zag chains and ring structures to three-dimensional spherical structures. Such an analysis
 provides us with the means to extract the general principles governing the ion transfer dynamics and 
 we gain some insight on the structural disorder caused by the quench of the barrier height. 
\end{abstract}

\maketitle
\section{Introduction}

Since their development many decades ago \cite{Paul1953,Penning1936,Dehmelt1990}, ion traps 
 have established themselves as a powerful tool in physics with applications ranging from  mass spectroscopy \cite{Daw1976,Mar1995}
to high precision tests for quantum electrodynamics \cite{Kreu2004,Stei2013,Gut2001}  and quantum information processing \cite{Cir1995,Wine1998,Kie2002}. 

The development of new trapping techniques, such as optical trapping  \cite{Scha2010} and the miniaturization of ion traps  on chip technologies 
 (microfabrication) \cite{Hugh2011, Wilp2012} opens new  possibilities for controlling the ions and accessing  physically interesting and 
yet unexplored trapping conditions. Consequently, the experimental and theoretical understanding of the behavior of 
 ions (both single species and mixtures) in different traps has
become the focus of many recent studies \cite{Haer2013,Smi2005,Gri2009,Champ2009}.

 A particular example is the study of Coulomb crystals. Using several cooling techniques such as Doppler cooling \cite{Win1979,Neu1978}, electromagnetically induced-transparency (EIT) \cite{Mor2000,Mor2003} or sympathetic cooling \cite{Mor2000}
 it is possible to reduce the kinetic energy of the trapped particles to the regime of $\mu$K 
 where the ions self-organize to the so-called Wigner crystals \cite{Wig_1934}. The structures of such crystals 
depend on the trap parameters and range from concentric rings (2D),
shells (3D) \cite{Boll1994,Dub1999, Boni2008} and string-of-disks configurations \cite{Drew2003} to two-component Coulomb bicrystals \cite{Drew2001}.
By tuning the  parameters of the trap potential, such as the amplitude  of the
AC and the amplitude and frequency of the DC potential in the case of a Paul trap  or the number of ions, Wigner crystals can undergo various transitions from one structure to another \cite{Bir1992,Dub1999}.
Special attention has been given in the literature to the case of the second order phase
transition from the linear to the zigzag chain of ions \cite{Mori20041,Mori20042,Balt20121}
which results in structures with \cite{Miel2012,Pyka2013,Ulm2013} or without \cite{Dub1993,Fish2008} topological defects (so-called kinks).
 In such a way the structural transitions of trapped ions serve among others as a playground for studying fundamental processes in physics, an example being that
of the Kibble-Zurek mechanism introduced originally in the field of cosmology \cite{Pyka2013}.

Given the wealth of effects resulting from a trapping of ions in an ordinary Paul trap  \cite{Paul1953} it is natural to ask for the effects 
stemming from a more involved trapping potential. Such a potential can be provided for ions through microfabrication, where for example
segmentation can be added to the standard Paul trap \cite{Schu2008,Tana2014} giving rise to a plethora of new possibilities
for trapping potentials \cite{kiel2002} like a double well 
 with tunable positions of minima used  for studying ion transport \cite{hub2008} or for splitting small ion crystals \cite{rust2014,kau2014,barr2004,rowe2002} .

Focusing on the case of the double-well trapping potential, the long-range inter- and intra-well interactions among the Wigner crystals occupying
each potential well give rise to a very complex non-equilibrium dynamics. In particular for a symmetric population of the two wells, we have recently shown  \cite{mine}
that a quench in the barrier height induces interesting nonlinear dynamics, involving both regular and irregular phases of motion. Here we extend this study to the 
case of an asymmetric population of the double well potential. This asymmetry results in an even richer dynamics involving ion transfers between the two wells, scattering processes
and crystal melting. A special focus is on the ion transfer mechanisms. We observe a non-smooth (step-like) dependence  of the time instant at which an arbitrary ion passes
above the barrier on the quench amplitude, relating to the oscillation frequencies of the ion closest to the barrier, as well as to the collective center-of-mass motion of both crystals.
After an ion transfer the smaller crystal melts due to the mass and energy excess a fact depicted nicely in the behaviour of the so-called Voronoi measure \cite{mine,Prep1985} of the crystal.
 
The paper is structured as follows. In section \ref{sec:setup} we present the general Hamiltonian of trapped ions in the effective potential of a Paul trap, introduce our setup
and describe its ground state configurations. Section \ref{sec:dynamics} is divided into two parts: we first present and analyze
 the ion transfer processes for crystalline structures  of different dimensionality and then discuss their effects on the order of the 
 two crystals, quantified by the Voronoi measure. At the end of section \ref{sec:dynamics} we comment briefly on the possibility of realizing our setup experimentally. Finally,
 in the last section \ref{sec:outlook} we summarize our results and give a short outlook for further possible investigations.

\section{Setup, Hamiltonian and ground state}
\label{sec:setup}
\subsection{General Hamiltonian}
We consider $N$ ions,  modeled as classical point particles with mass $m$ and charge $Q$, confined  along the radial direction ($x,y$) to a linear quadrupole Paul trap and
along the axial  direction $z$ to a double well potential (segmented trap)\footnote{The combination of the linear quadrupole trap potential and a double well potential is not
allowed by the Laplace equation. Therefore it would be necessary
to compensate for it by a complex form of the dc part in the radial direction. In recent experiments with segmented Paul traps (e.g. \cite{Brow2011,Harl2011}) 
this combination was realized, giving rise to a potential similar to the one we have chosen here as well as to a Mexican hat like potential.}.

 The general expression for the radially confining potential $\Phi(x,y,t)$  (in the ($x,y$)-plane) reads
 
\begin {equation} 
\Phi(x,y,t) =\frac{U_{\rm dc}}{2}(cx^2+cy^2)+\frac{U_{rf}}{2}\cos{(\omega_{\rm rf} t)}(cx^2-cy^2). 
\end {equation}
where $U_{\rm dc}$ and $U_{\rm rf}$ are the applied constant and radiofrequency (rf) voltages  respectively; $\omega_{\rm rf}$ denotes the (radio)frequency and $c$ is a parameter 
specifying the geometry of the trap.

The ion dynamics in the radial rf trap is composed of  a fast motion, the so-called micro-motion, and a comparatively slow average motion ruled by an effective harmonic potential
\cite{Ger1992} $ V(x,y)=\frac{m}{2}(\omega_x^2 x^2+\omega_y^2 y^2)$.
Here, $\omega_{x}=\frac{\omega_{rf}}{2}\sqrt{a-q^2/2}$ and $\omega_{y}=\frac{\omega_{rf}}{2}\sqrt{a+q^2/2}$ are the effective trapping frequencies with 
$a=\frac {4QU_{dc}}{m \omega^2_{rf}}c$ and $q=\frac{2QU_{rf}}{m\omega_{rf}^2}c$ being dimensionless parameters. In the following 
we will neglect the fast micro-motion, for reasons of simplicity, being less  relevant for the averaged system dynamics.

\begin{figure}[h!]
 \begin{tikzpicture}[node distance=5em, scale=0.4]
   \node at (0,0){\includegraphics[bb=0 0 842 334,scale=0.4,keepaspectratio=true]{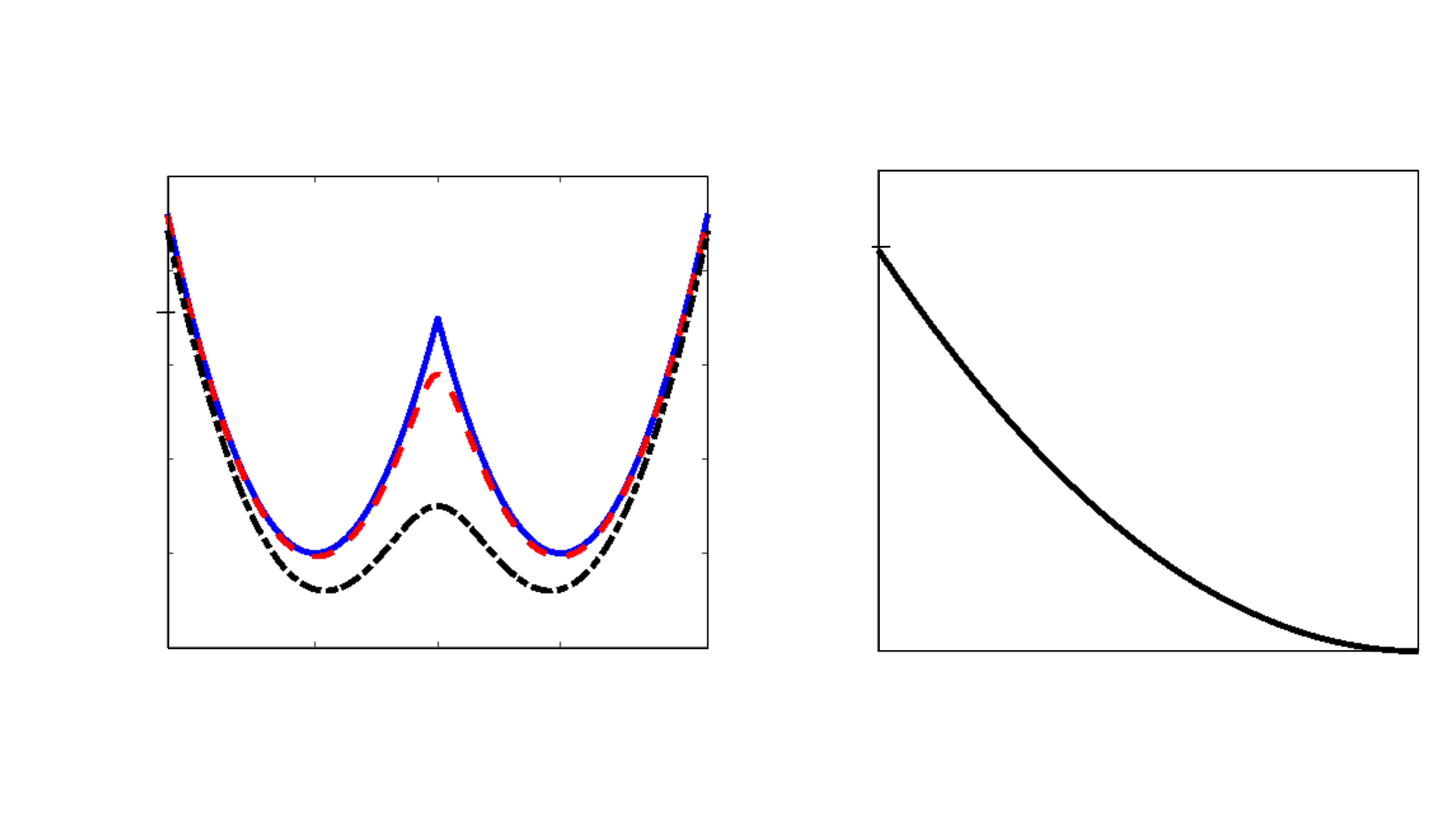}};
   \node at (-10.2,7.6) {\large (a)};
   \node at (5.6,7.7) {\large (b)};
   \node at(-7.8,-2.7){$-z_0$};
   \node at(-2.2,-2.7){$z_0$};
   \node at(-5,-2.7){$0$};
   \node at(4.6,-2.7){$0$};
   \node at(17.,-2.7){$z_0^2$};
   \node at(-5,-4.2){\large axial direction ($z$)};
   \node at (-5,-5.7){(units in $\nicefrac{1}{K}$)};
    \node at(11,-4.2){\large $C$ };
    \node at(11,-5.7){(units in $\nicefrac{1}{(K^2\omega_z^2)}$)};
   \node at(-11.8,0){$0$};
   \node[rotate=90] at (-15,3){\large $V_d(z)$};
   \node[rotate=90] at (-13.5,3){(units in $\nicefrac{K^2}{(m\omega_z^2)}$)};
   \node[rotate=90] at (2.2,3){\large $B(C)$ };
    \node[rotate=90] at (3.5,2.5){(units in $\nicefrac{K^2}{(m\omega_z^2)}$)};
   \node at(-12.2,5.5){$\nicefrac{z_0^2}{2}$};
   \node at(3.6,7.){$\nicefrac{z_0^2}{2}$};
 
  \end{tikzpicture}
  \caption{(a) The double well trapping potential in axial direction $V_d(z)$ for different values of $C$ (blue solid line $C=0.02$, red dashed line $C=3$, black dashed-dotted line  $C=10$)
and (b) the barrier height $B$ as a function of $C$.}
\label{Fig.:barrier}
\end{figure}
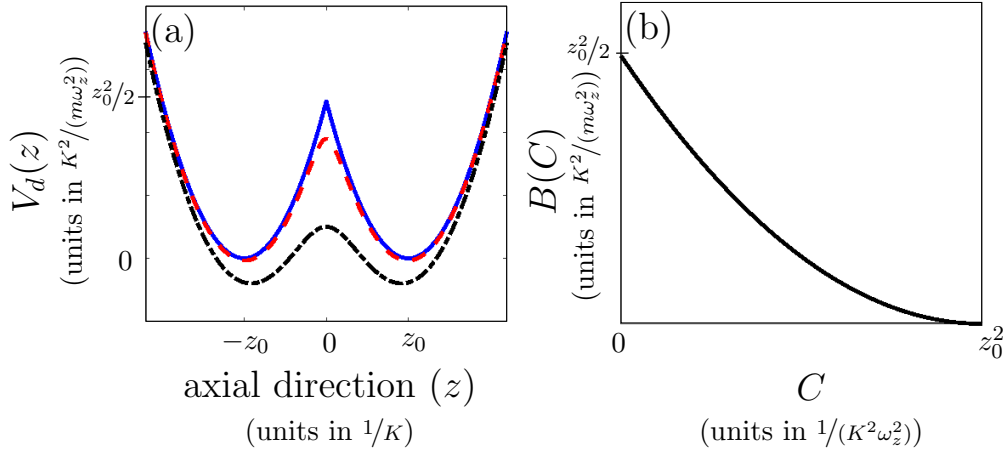

For the confinement in the $z$-direction we assume the following phenomenological double-well potential \cite{Stre2006}:

\begin{equation}
V_{\rm d}(z) =\frac{m}{2}\omega_z^2z_0^2+\frac{m}{2}\omega_z^2z^2-\frac{m}{2}\sqrt{4C^2+4\omega_z^4z^2z_0^2} \label{axpot1} 
\end{equation}
with wells centered at $\approx\pm z_0$ and separated by a barrier of height
\begin{equation}
 B=\frac{m}{2}(\omega_z^2z_0^2+\frac{C^2}{\omega_z^2z_0^2})-mC  \qquad\text{with} \quad C\in (0\, ,\omega_z^2z_0^2]
\end{equation}
illustrated in Fig.~\ref{Fig.:barrier} (a),(b).
It can be shown that such a potential leads to  individual approximately harmonic wells centered at $\approx\pm z_0$, up to terms proportional to $C^2$.


Under the aforementioned assumptions, the total Hamiltonian of our system, including the radial $V(x,y)$ and the axial $V_{\rm d}(z)$ trapping potentials as well as
 the Coulomb interactions among the ions reads

\begin{align} H(\{ {\bf r}_i,{\bf p}_i\})=&\sum_{i=1}^n\frac{{{\bf p}_i}^2}{2m}+\sum_{i=1}^n \left[V_{\rm d}(z_i)+V(x_i,y_i)\right]\nonumber \\
&+\sum_{i=1,j<i}^n \frac{Q^2}{4\pi\epsilon_0 r_{ij}} \label{hami1}
\end{align}

with ${\bf r}_i=(x_i,y_i,z_i)$,
$r_{ij}=\sqrt{(x_i-x_j)^2+(y_i-y_j)^2+(z_i-z_j)^2}$ and ${\bf p}=(p_x,p_y,p_z)$.
We arrive then at the dimensionless Hamiltonian $H^*$ by introducing rescaled time $t_u=1/\omega_z$ and space  $x_u=K\equiv[Q^2/(4\pi\epsilon_0m\omega_z^2)]^{1/3}$ units and defining

\begin{align} 
&t^*=\omega_z t;\; x^*=\frac{x}{K};\; y^*=\frac{y}{K};\; z^*=\frac{z}{K};\;z_0^*=\frac{z_0}{K};\;r_{ij}^*=\frac{r_{ij}}{K};\nonumber \\ 
&C^*=\frac{C}{K^2\omega_z^2};\; \alpha=\frac{\omega_x}{\omega_z};\;  \beta=\frac{\omega_y}{\omega_z}.
\label{eq:units}
\end{align}

Note that in the following we omit the star for simplicity and we present all our results in these dimensionless units.

\subsection{ Specific setup parameters and ground state configurations}
Having the dimensionless Hamiltonian we proceed to find its ground state (GS) configuration. Finding the global minimum of a many-ion 
potential is generally a highly nontrivial task. In our case, however, and for large enough values of the potential barrier $B$ (low values of $C$, Fig.~\ref{Fig.:barrier} (b))
it turns out that we can find the GS configuration by using a root-finding algorithm \cite{gsl} given as an initial guess the GS of the ions in two
individual harmonic wells (approximating the double well potential) which is known in the literature \cite{Dub1993,Drew2001,Drew2003PhysB,Bedan1993,Ludw2005}.

Using this GS configuration as the initial state of the system, we then perform  a quench in the barrier height. In order to analyze the non-equilibrium dynamics
of the system following the quench we integrate the resulting Newtonian equations of motion (EOM) employing an implicit Gaussian 4th order Runge-Kutta algorithm \cite{gsl}.
\begin{figure}[h!]
\vspace{-3cm}
 \begin{tikzpicture}[node distance=5em, scale=0.45]
   \node at (0,0){\includegraphics[bb=0 0 1326 854,scale=0.45,keepaspectratio=true]{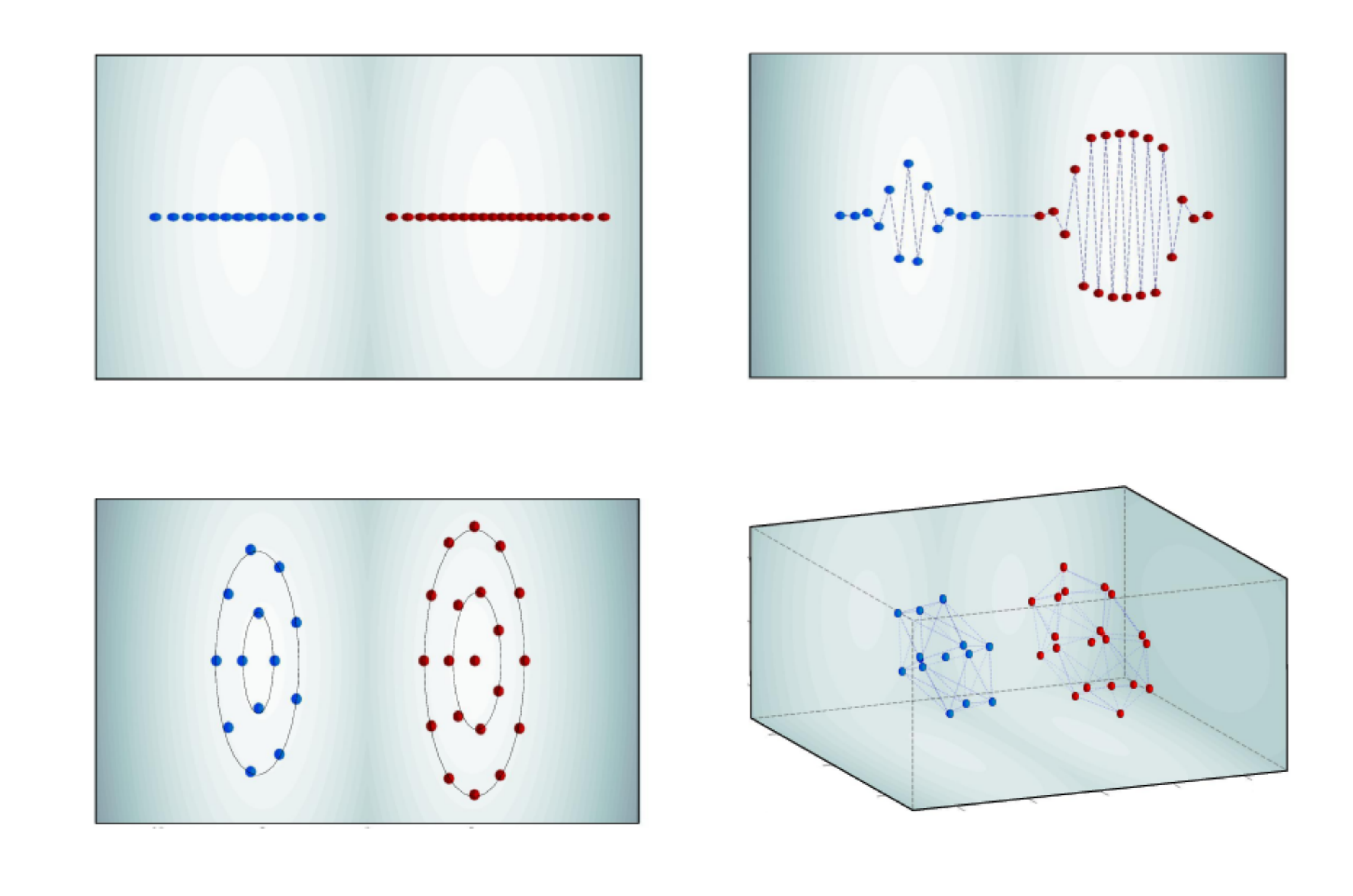}};
   \node at (-22.,5.7) {\large (a)};
   \node at (-6.6,5.7) {\large (b)};
   \node at (-22.2,-5.7) {\large (c)};
   \node at (-6.6,-5.7) {\large (d)};
   \node[rotate=90] at(-22.7,1.5){\large radial};
   \node[rotate=90] at(-6.6,-8.5){\large radial};
   \node[rotate=90] at(-22.7,-9.5){\large radial};
   \node[rotate=-33] at(-4.57,-12.7){\large radial};
   \node[rotate=90] at(-6.7,1.5){\large radial};
   \node at(-14.5,-14.7){\large axial};
 \node at(-14.5,-4.2){\large axial};
 \node at(1.4,-4.2){\large axial};
 \node[rotate=8] at(3.8,-13.9){\large axial};
 \node at(-17.6,-3.3){$-z_0$};
 \node at(-11.4,-3.3){$z_0$};
 \node at(-17.3,-13.8){$-z_0$};
 \node at(-11.7,-13.8){$z_0$};
 \node at(-1.2,-3.3){$-z_0$};
\node at(4,-3.3){$z_0$};
\node at(1.5,-13.3){$-z_0$};
 \node at(5.5,-13){$z_0$};
 \node at(-14.5,-3.3){$0$};
 \node at(1.4,-3.3){$0$};
 \node at(3.8,-13.1){$0$};
 \node at(-14.5,-13.8){$0$};
\node at(-21.5,1.6){$0$};
 \node at(-21.5,-9.5){$0$};
\node at(-5.7,1.6){$0$};
 \node at(-5.7,-8.4){$0$};
  \node at(-3.9,-12.){$0$};
\end{tikzpicture}
  \caption{Initial ion ground state configurations for different potential parameters: (a) linear chains for $\alpha=100,\beta=100$,
(b) zig-zag chains for $\alpha=5.6,\beta=8$, (c) circles for $\alpha=1,\beta=8$, (d) spheres for $\alpha=1,\beta=1$.
All configurations and their dynamics are calculated in three dimensions  with their dimensionality being restricted only by the frequency ratios $\alpha$ and $\beta$.
}
\label{Fig.initial}
\end{figure}
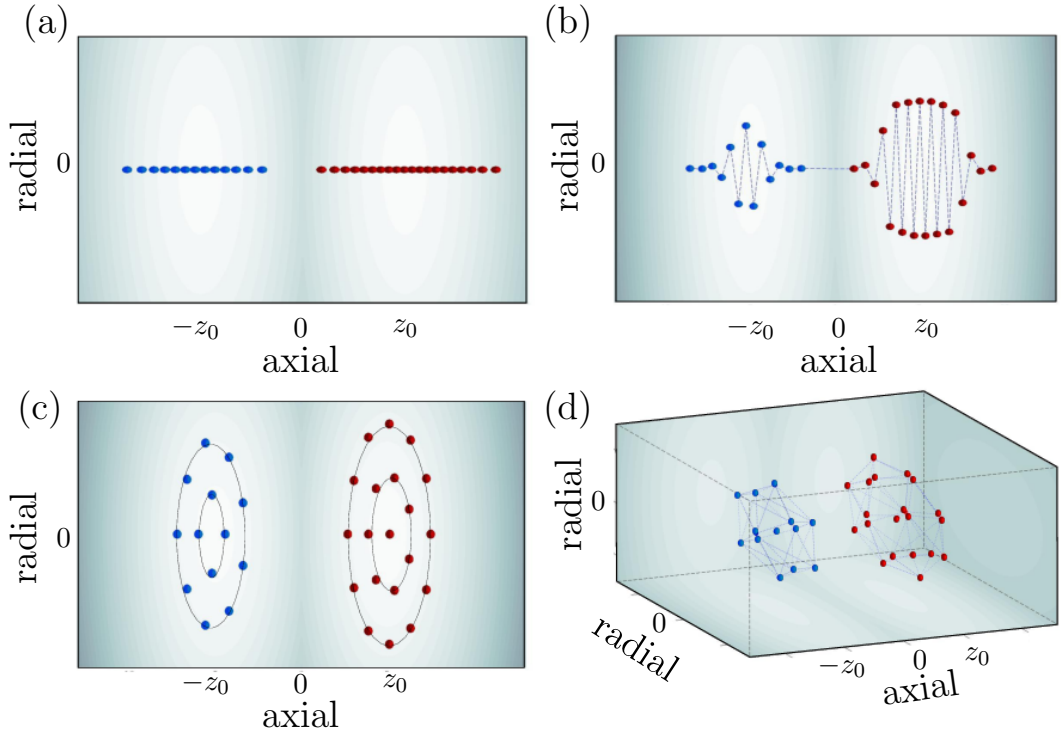

 In this paper we study a system of $N=33$ ions confined in a combined trap with a harmonic potential in the radial direction and a double-well potential 
 in the axial direction, described by eq. (\ref{hami1}). Initially the barrier height in the axial direction has a large value corresponding to  $C_{i}=0.02$
 (eq. (\ref{axpot1}), fig.~ \ref{Fig.:barrier}(a)) leading to well-separated potential wells which are asymmetrically filled with ions, i.e.
 the right potential well contains $N_R=20$ ions, whereas  the left potential well contains $N_L=13<N_R$ ions.
 Regarding the radial confinement, we examine here four qualitatively different cases corresponding to different aspect ratios 
 $\alpha=\frac{\omega_x}{\omega_z}$ and $\beta=\frac{\omega_y}{\omega_z}$ and allowing for GS configurations of different dimensionality, ranging from linear to  zig-zag,
 circular and  spherical crystals (Fig.~\ref{Fig.initial}). In order to take into account also the effect of a finite (low) temperature,
 we add a small random initial velocity to the ions leading to small oscillations around their GS positions. 

\section{Ion dynamics}
\label{sec:dynamics}
 Having described our setup allowing for a variety of equilibrium configurations (see Fig.~\ref{Fig.initial}), we now proceed to investigate the resulting dynamics, induced
here by a change of the barrier height. In particular
we are interested in the transfer dynamics of the ions following a sudden quench of the barrier height which is 
characterized by the parameter $C$, i.e. a quench of the axial potential $V_{\rm d}$ from $C=C_{i}$ to  $C=C_f$. 
We investigate therefore in the following the non-equilibrium dynamics of the Coulomb crystals in different confining potentials 
(Fig.~\ref{Fig.initial}) as a function of $C_f$.

With all the necessary information about the properties of our system at hand, let us now briefly introduce some of the basic features of its non-equilibrium dynamics.
After a sudden quench of the barrier height, due to the energy excess, the ions constituting the Coulomb crystals start to move.
Their dynamics is complex, involving among others a rather regular center of mass (CM) motion of the crystals, shock waves, multiple scattering processes and transfer of ions
over the barrier caused by the asymmetric population of the two potential wells. Especially for the cases of the
 circular and spherical GS configurations also rotations and reordering of the crystals occur.  

 We first proceed in analyzing the ion transfer process between the two potential wells and then discuss its effects on the structure of the two Coulomb crystals.

\subsection{Ion transfer}

Among the features characterizing the ion transfer following a quench of the barrier height we are particularly
interested in two aspects, to be analyzed below: 
the time instant at which an arbitrary ion passes above the barrier for the first time and the number of times each ion in the Coulomb crystal travels back and forth 
between the two wells.

\begin{figure}[h!]
\hspace{-2cm}
\begin{tikzpicture}[node distance=5em, scale=0.45]
   \node at (0,0){
 \includegraphics[bb=0 0 1137 793,scale=0.45,keepaspectratio=true]{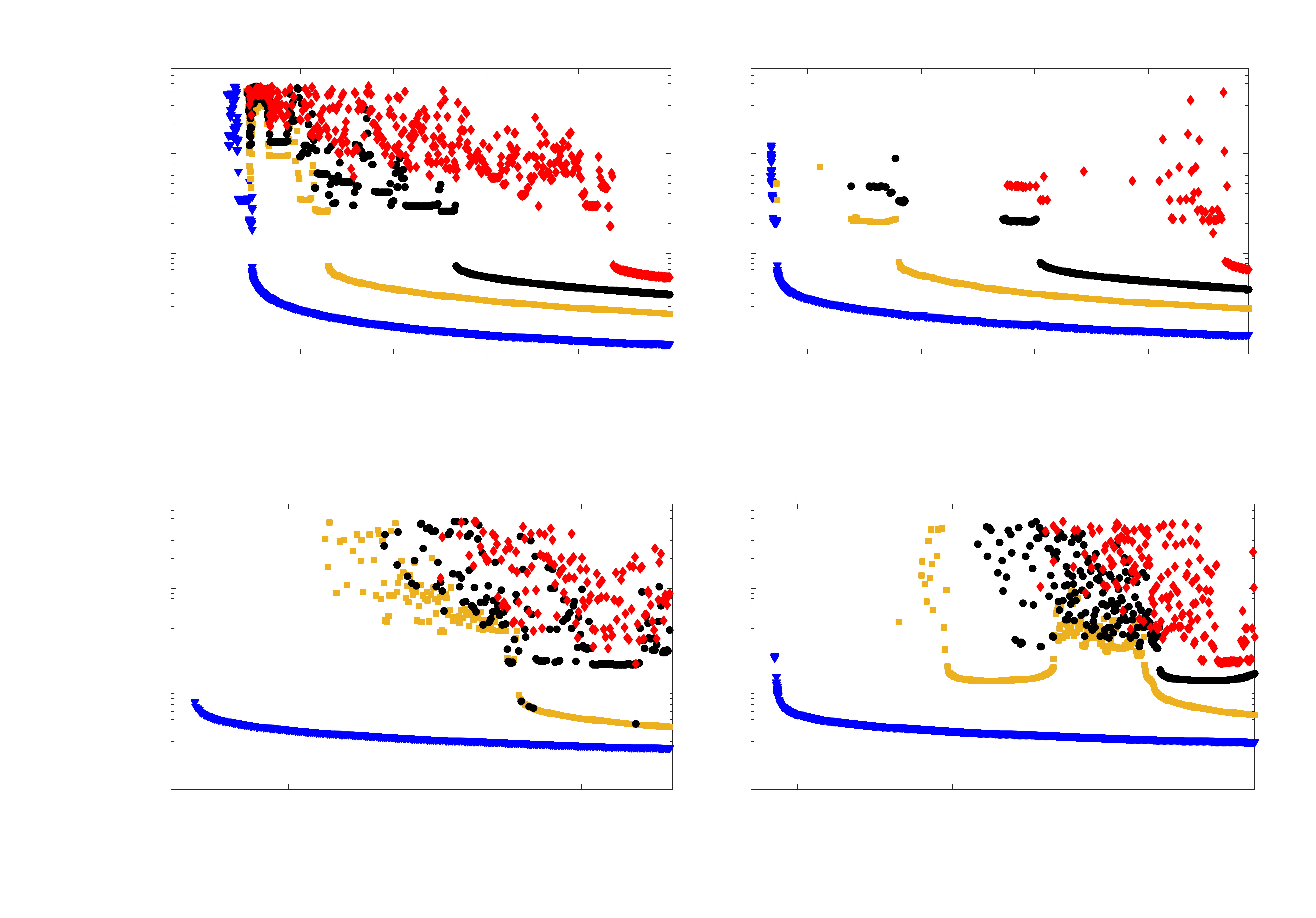}};
 \node[rotate=90] at(-17,7.6){\large time (units in $\nicefrac{1}{\omega_z}$)}; 
 \node[rotate=90] at(-17,-5.7){\large time (units in $\nicefrac{1}{\omega_z}$)};
 \node at(-8.,-11.9){\large $C_f$ (units in $\nicefrac{1}{(K^2\omega_z^2)}$)};
 \node at(11,-11.9){\large $C_f$ (units in $\nicefrac{1}{(K^2\omega_z^2)}$)};

   \node at(-14.,11.) {\large (a)};
   \node at(3.8,11.) {\large (b)};
   \node at(-14,-2.2) {\large (c)};
   \node at(3.8,-2.2) {\large (d)};
 \node at(-15.7,3.15){$10^0$};
\node at(-15.7,6.3){$10^1$};
 \node at(-15.7,9.5){$10^2$};
 
 \node at(2,3.15){$10^0$};
\node at(2,6.3){$10^1$};
 \node at(2,9.5){$10^2$};
 
 \node at(2,-10.15){$10^0$};
\node at(2,-7.){$10^1$};
 \node at(2,-3.8){$10^2$};
 
 \node at(-15.7,-10.15){$10^0$};
\node at(-15.7,-7.){$10^1$};
 \node at(-15.7,-3.8){$10^2$};
 \node at(-13.7,2.3){$2.5$};
\node at(-10.9,2.3){$3$};
 \node at(-7.9,2.3){$3.5$};
 \node at(-5.3,2.3){$4$};
 \node at(-2.5,2.3){$4.5$};
 \node at(.5,2.3){$5$};
 \node at(4.6,2.3){$3.5$};
 \node at(8.,2.3){$4$};
 \node at(11.5,2.3){$4.5$};
 \node at(15,2.3){$5$};
 \node at(-11.3,-10.8){$9$};
 \node at(-6.8,-10.8){$10$};
 \node at(-2.3,-10.8){$11$};
 \node at(4.2,-10.8){$10$};
 \node at(9,-10.8){$11$};
 \node at(13.8,-10.8){$12$};
 \end{tikzpicture}
  \caption{Time instant of the first transfer of the first four traveling ions (order: first- blue triangle, second- yellow squares, third- black diamonds, fourth- red dots)
  as a function of $C_f$ for a) linear chains b) zig-zag chains c) circles d) spheres. 
   For lower  $C_f$ values than the ones depicted in the figures there is no ion transfer as explicitly shown in (a).}
\label{Fig:point_in_time}
\end{figure}
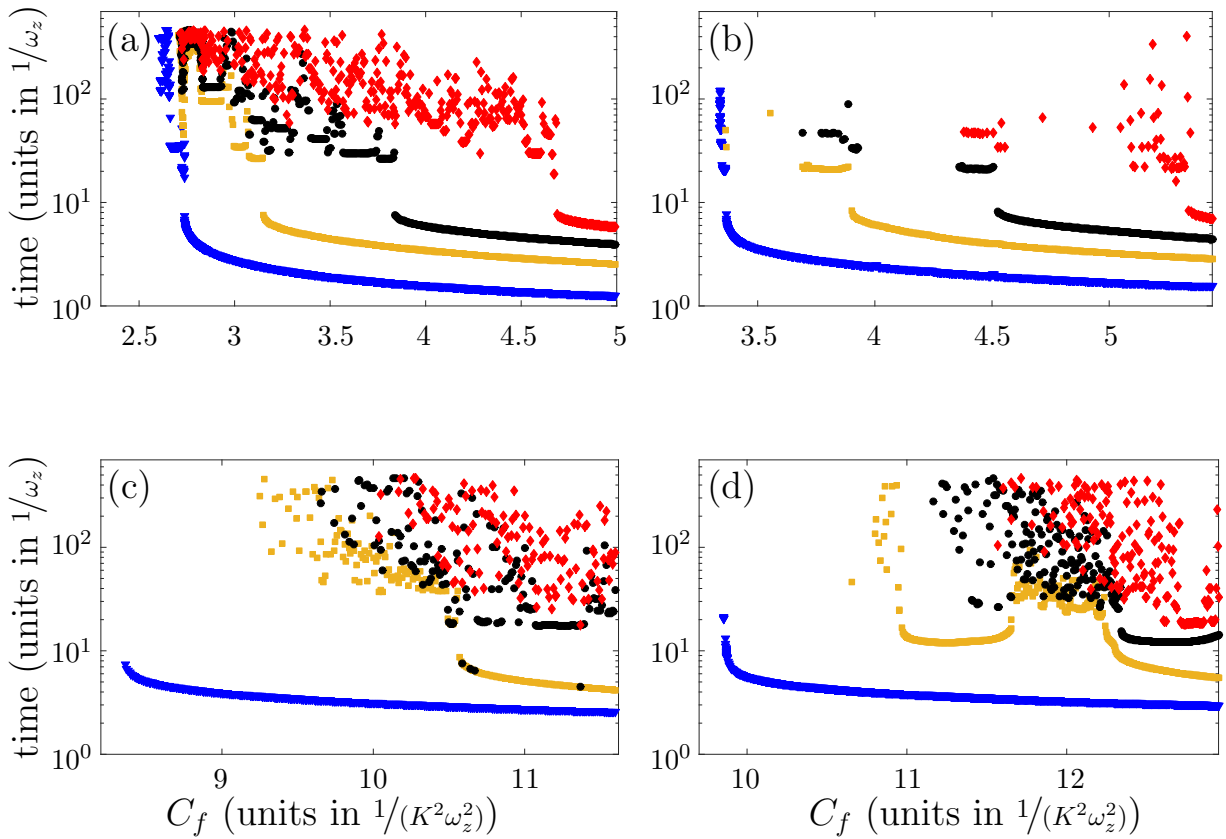

Our results for the time instant of the first transfer of an arbitrary ion leaving the large crystal as a function of the final quench value 
$C_f$ are presented in Fig. \ref{Fig:point_in_time} for the different trapping geometries depicted in Fig. \ref{Fig.initial}. For the facility of inspection we show only the first four traveling ions. In all the cases one can distinguish 
between two qualitatively different regions.
The first one occupying the upper part of the plots (large transfer times $t>10$) shows a step-like behaviour (with varying $C_f$) consisting of small smooth regions separated by gaps, whereas the lower part
(small transfer times $t<10$) exhibits a continuous behaviour. We see that depending on the value of $C_f$ 
more than one ion can be transferred. Each individual ion transfer follows the same qualitative behaviour with respect to its time scale. For small quench amplitudes and subsequently large transfer times the time instant of the first transfer of each ion
has a step-like character, whereas beyond a certain quench amplitude the transfer time continuously decreases as a function of the final value $C_f$.

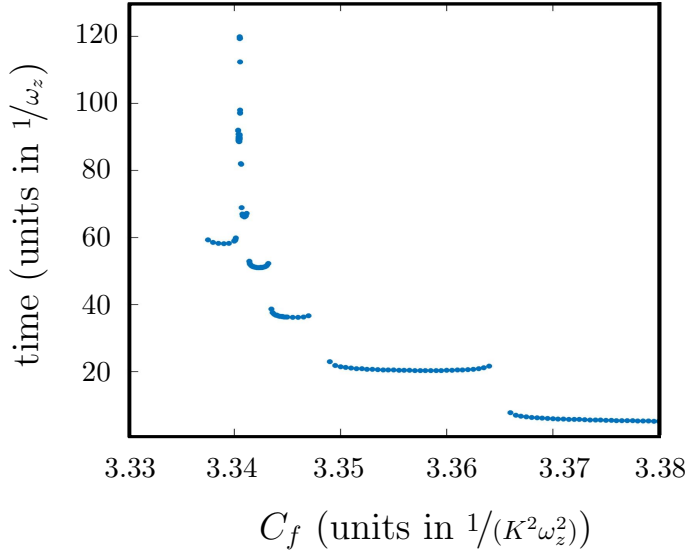
\begin{figure}[h!]
 \begin{tikzpicture}[node distance=5em, scale=0.15]
    \node at (0,0){
 \includegraphics[bb=0 0 3188 1753,scale=0.15,keepaspectratio=true]{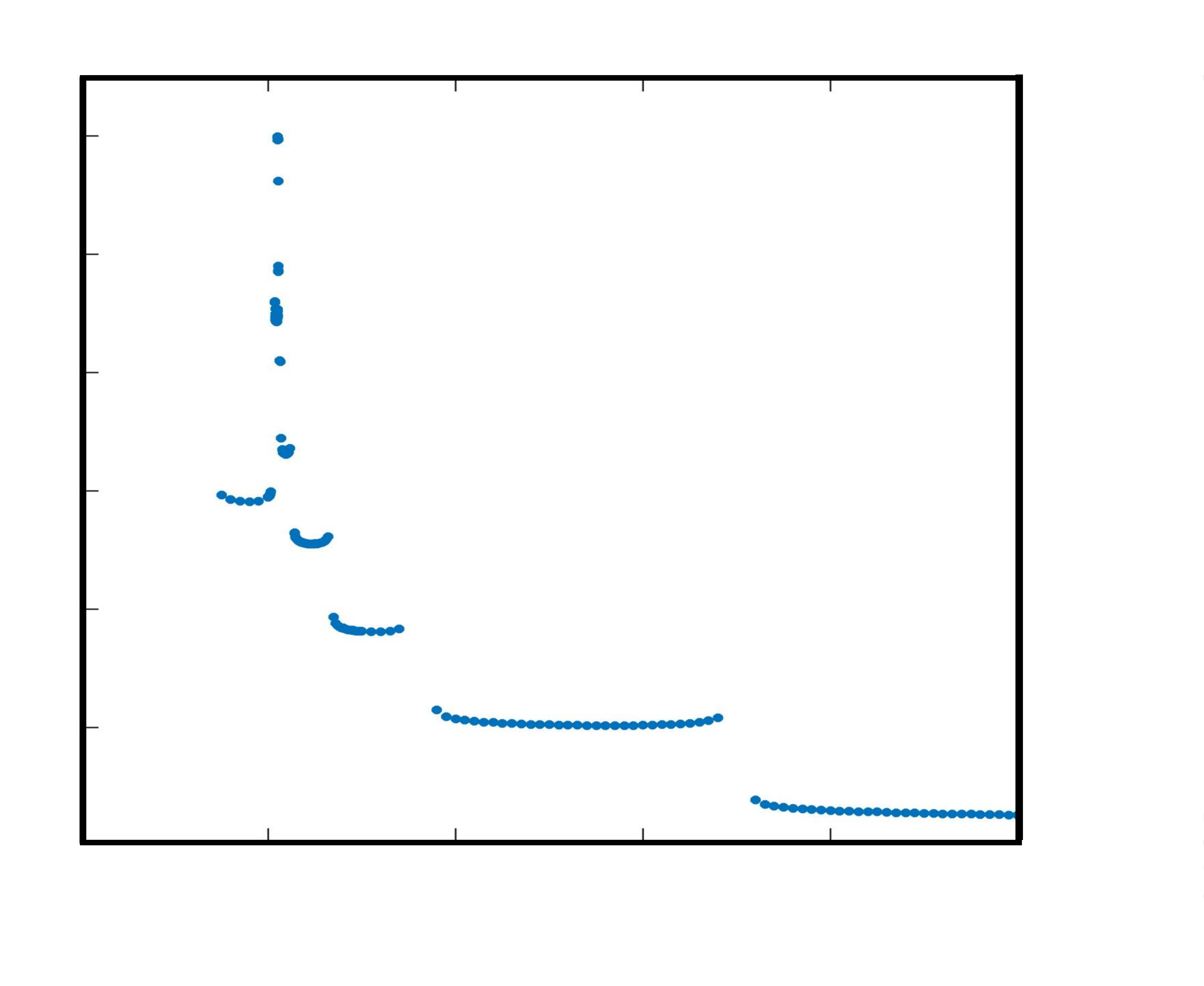}};
   \node[rotate=90] at(-61,-4) {\large time (units in $\nicefrac{1}{\omega_z}$)};
 \node at(-25.8,-32) {\large $C_f$ (units in $\nicefrac{1}{(K^2\omega_z^2)}$)};
 \node at(-52,-26.5) {$3.33$};
 \node at(-43,-26.5) {$3.34$};
 \node at(-33.5,-26.5) {$3.35$};
 \node at(-23.5,-26.5) {$3.36$};
 \node at(-13.5,-26.5) {$3.37$};
 \node at(-5,-26.5) {$3.38$};

 \node at(-55,-18) {$20$};
 \node at(-55,-12) {$40$};
\node at(-55,-6) {$60$};
 \node at(-55,0) {$80$};
 \node at(-55,6) {$100$};
 \node at(-55,12) {$120$};
\end{tikzpicture}\caption{Time instant of the first transfer of the first traveling ion as a function of $C_f$ for the case of the zig-zag chain (zoom of Fig.~\ref{Fig:point_in_time}(b))}
 \label{Fig.:begin_zigzag}
\end{figure}

As a characteristic example of this behaviour, without loss of generality in the remaining part of this subsection, we focus on the case of the zig-zag configuration (Fig. \ref{Fig:point_in_time} (b))
and consider only the transfer of the first ion (Fig. \ref{Fig.:begin_zigzag}). 
 The first necessary condition for the transfer of an ion above the barrier is obviously that this ion lies close to the barrier.
To examine therefore the possibility of transfer it is instructive to analyze the dynamics of the ion of the bigger crystal lying 
closest to the barrier, which we will refer to in the following as the innermost ion. Such an analysis can be facilitated by examining a case in which ion transfer
although energetically possible does not occur, i.e. corresponding to a quench value inside a gap of Fig. \ref{Fig.:begin_zigzag}.
A case satisfying these criteria and consequently allowing for an analysis of the long-time dynamics of the innermost ion 
without interruptions by ion transfer processes is that of a final quench value  $C_f=3.3403$. 

\begin{figure}[h!]
\vspace{-10cm}
 \begin{tikzpicture}[node distance=5em, scale=0.35]
     \node at (0,0){\includegraphics[bb=0 0 2550 1412,scale=0.35,keepaspectratio=true]{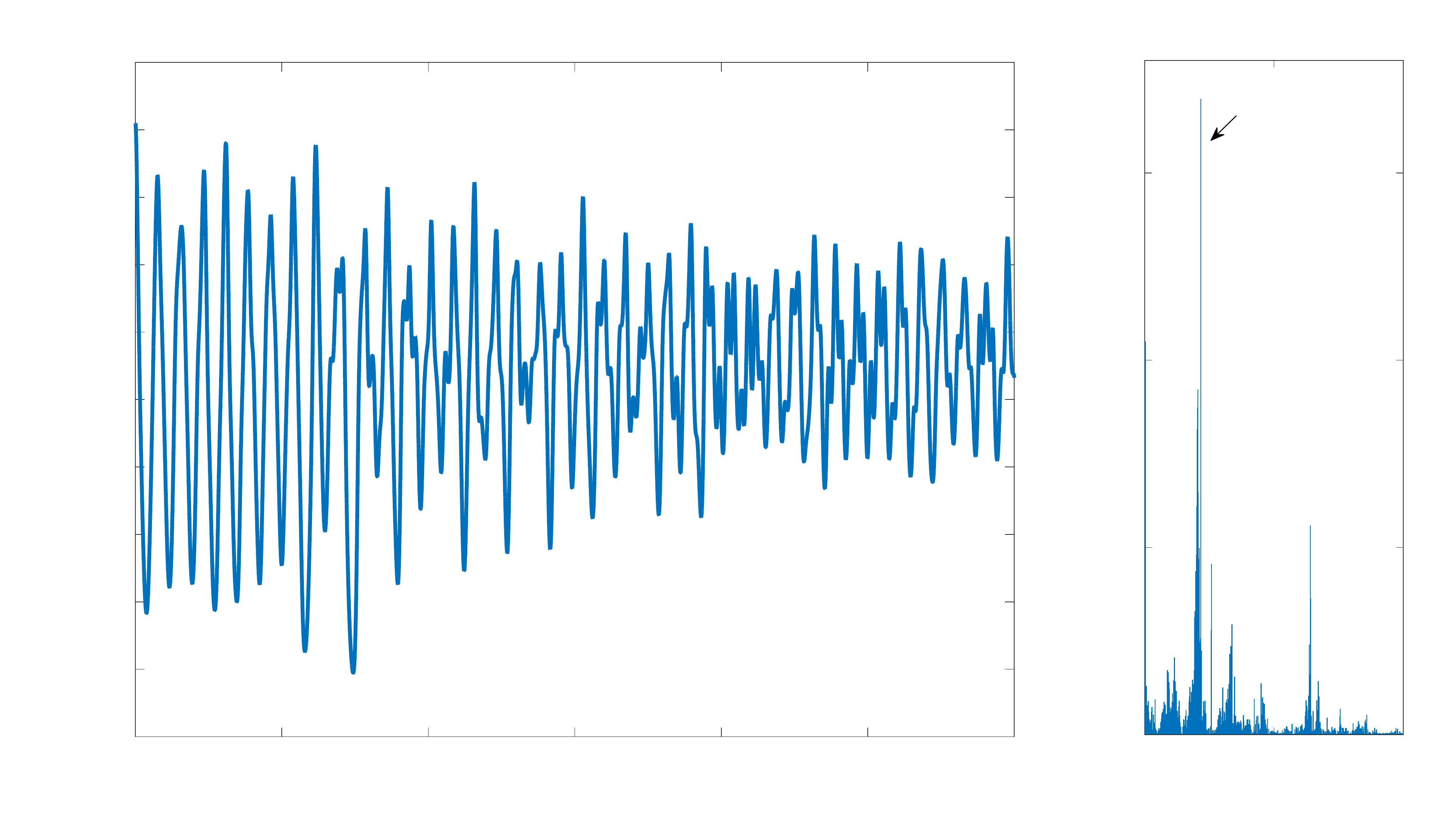}};
   \node at (-40.1,-4.7) {\large (a)};
   \node at (-12.5,-4.7) {\large (b)};
 \node[rotate=90] at(-44,-13) {\large axial coordinate (units in $\nicefrac{1}{K}$)};
 \node at(-28,-26) {\large time (units in $\nicefrac{1}{\omega_z}$)};
 \node at(-9,-26) {\large $\omega$ (units in $2\pi \omega_z$)};
\node[rotate=90] at(-16,-13) {\large amplitude};
 \node at(-41.5,-23.5) {$0$};
 \node at(-33.2,-23.5) {$100$};
\node at(-25.,-23.5) {$200$};
 \node at(-16.7,-23.5) {$300$};
 \node at(-42.4,-16.8) {$0.5$};
 \node at(-42.4,-7.7) {$1$};
 \node at(-14,-23.5) {$0$};
 \node at(-10,-23.5) {$2$};
 \node at(-6.2,-23.5) {$4$};
\node at(-14.5,-17) {$5$};
 \node at(-14.5,-12.) {$10$};
 \node at(-14.5,-7) {$15$};
\node at(-8.85,-4.75) {$\omega=0.868$};
  \end{tikzpicture}\caption{The axial motion of the innermost ion of the zig zag configuration and  its Fourier spectrum for $C_f=3.3403$}
\label{Fig.:kompl_66}
\end{figure}
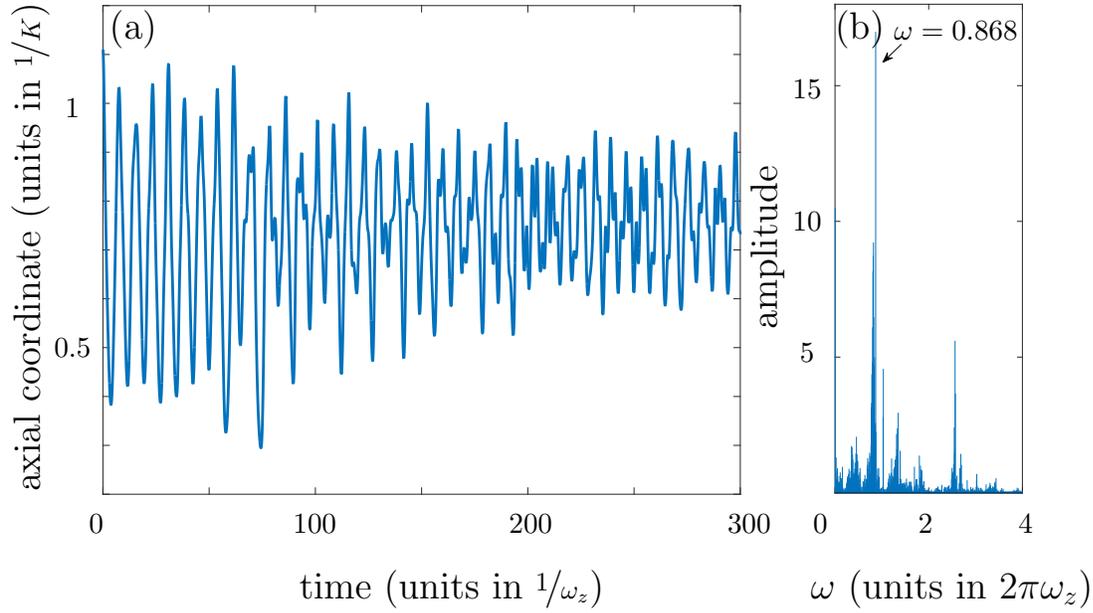
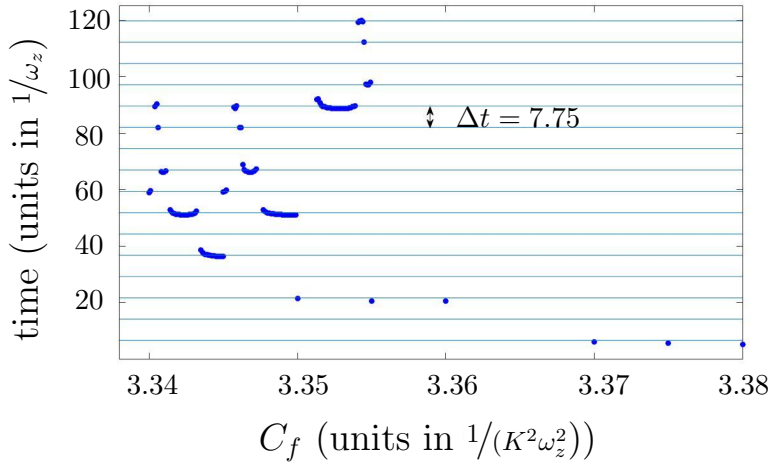
\begin{figure}[h!]
 \begin{tikzpicture}[node distance=5em, scale=0.30]
   \node at (0,0){
 \includegraphics[bb=0 0 1261 740,scale=0.30,keepaspectratio=true]{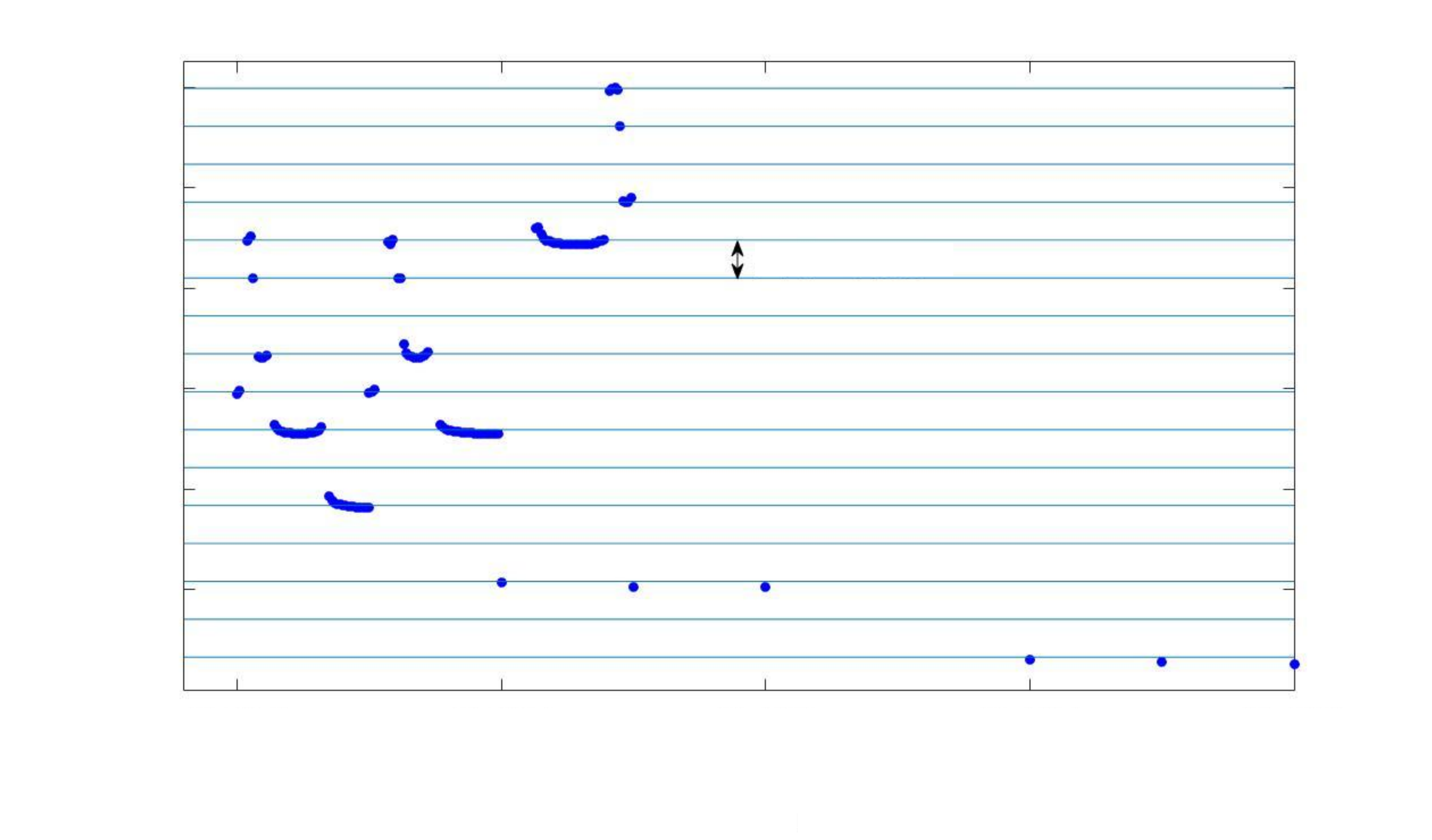}};
 \node[rotate=90] at(-21.7,-1.8) {\large time (units in $\nicefrac{1}{\omega_z}$)};
 \node at(-4,-13) {\large $C_f$ (units in $\nicefrac{1}{(K^2\omega_z^2})$)};
 \node at(0,1.4){$\Delta t=7.75$};
 \node at(-19,-6.7) {$20$};
\node at(-19,-4.3) {$40$};
 \node at(-19,-1.9) {$60$};
\node at(-19,.55) {$80$};
\node at(-19,3.1) {$100$};
 \node at(-19,5.7) {$120$};
 \node at(-16.2,-10.5) {$3.34$};
\node at(-9.5,-10.5) {$3.35$};
 \node at(-2.9,-10.5) {$3.36$};
 \node at(3.8,-10.5) {$3.37$};
 \node at(10,-10.5) {$3.38$};

 \end{tikzpicture}
 \caption{ The time instant for the first transfer of the innermost ion as a function of $C_f$ for the zig-zag configuration. The horizontal lines mark 
 the multiples of the period $T$ of the innermost ion.}
 \label{Fig.:period}
\end{figure}

The time evolution of the axial position of the innermost ion of the large crystal for $C_f=3.3403$, as well as its Fourier spectrum 
are shown in Fig. \ref{Fig.:kompl_66}. The motion appears to be oscillatory and quite regular with one dominant frequency, a fact supported also by the Fourier spectrum 
which shows essentially the contribution of three frequencies, with one of them possessing a dominant role (largest amplitude).
In order to test whether this frequency is generic for our system, we have investigated the Fourier spectra of the innermost ion motion for four different
values of $C_f$ which do not lead to transfer.
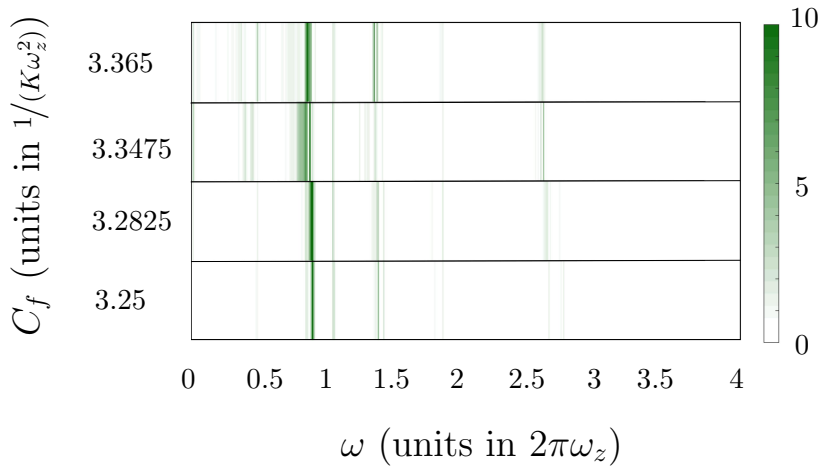
\begin{figure}[h!]
\vspace{-11cm}
 \begin{tikzpicture}[node distance=5em, scale=0.3]
     \node at (0,0){\includegraphics[bb=0 0 3188 1753,scale=0.3,keepaspectratio=true]{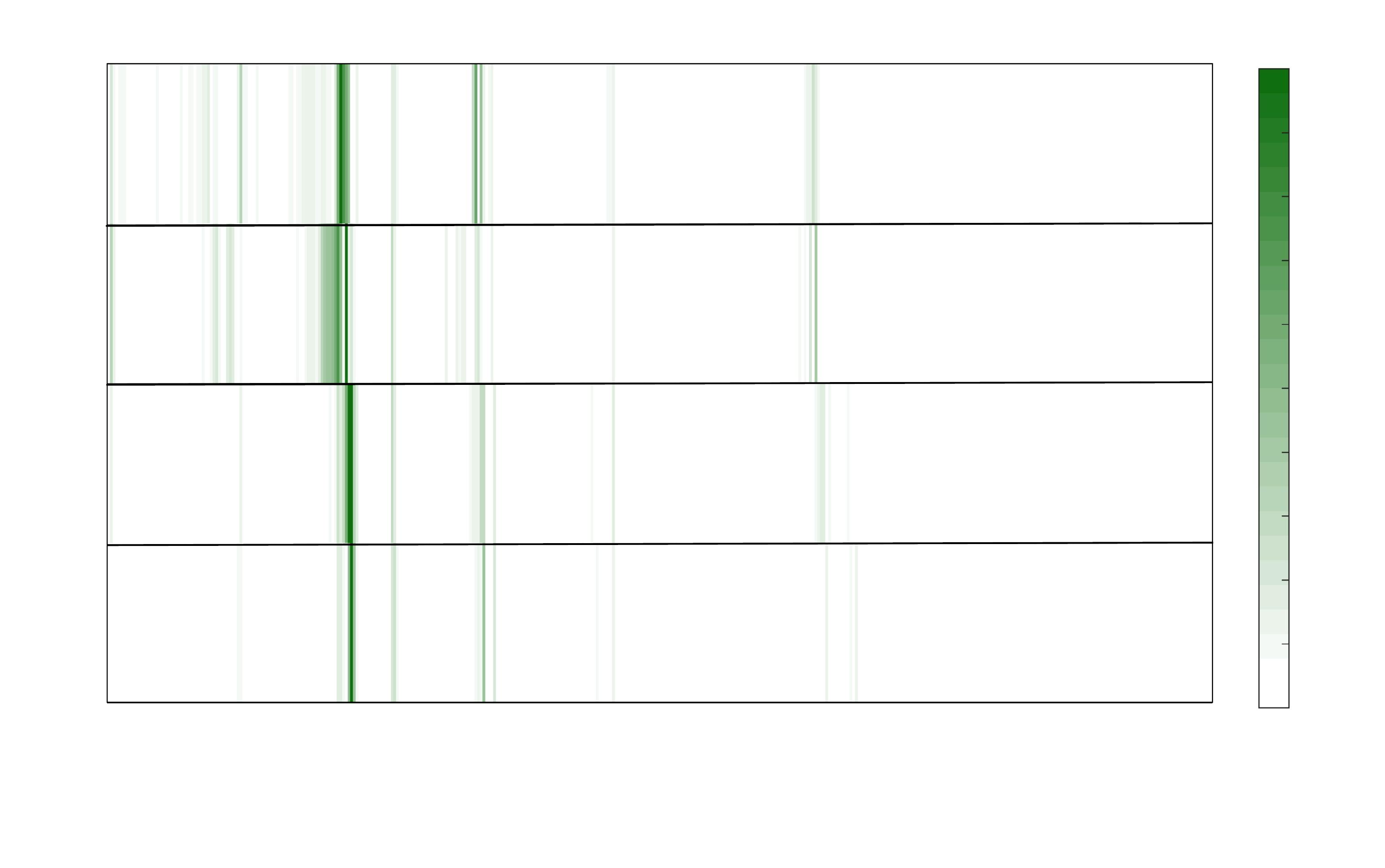}};
\node[rotate=90] at(-61,-20) {\large $C_f$ (units in $\nicefrac{1}{(K\omega_z^2)}$)};
 \node at(-41,-32) {\large $\omega$ (units in $2\pi \omega_z$)};
 \node at(-54,-29){$0$};
\node at(-26.8,-20.5){$5$};
 \node at(-26.7,-13){$10$};
 \node at(-26.8,-27.5){$0$};
 \node at(-50.6,-29){$0.5$};
 \node at(-47.9,-29){$1$};
 \node at(-45.,-29){$1.5$};
 \node at(-42.1,-29){$2$};
\node at(-39.,-29){$2.5$};
 \node at(-36.0,-29){$3$};
 \node at(-33.3,-29){$3.5$};
 \node at(-29.7,-29){$4$};
 \node at(-57,-25.5){$3.25$};
 \node at(-56.5,-22){$3.2825$};
 \node at(-56.5,-18.8){$3.3475$};
\node at(-57,-15){$3.365$};
 \end{tikzpicture}
 \caption{Comparison of the Fourier spectra of the innermost ion motion for 4 different values of $C_f$. The colour depicts their 
 amplitude.}
 \label{Fig:Vergleich}
\end{figure}
The results are depicted in Fig.~\ref{Fig:Vergleich}. Obviously in all the cases there is one predominant frequency in the range of $\omega\approx 0.84$ to $0.88$,
which in turn yields a period $T$ ranging approximately from $7.1$ to $7.4$.  A comparison of this period with the time between the subsequent steps in Fig. \ref{Fig.:begin_zigzag}
leads to the conclusion that the latter is equal to one or two times the period $T$ (Fig. \ref{Fig.:period}) within a range of $3.5\%$.
This fact explains the existence of the steps in the ion transfer process as a direct consequence of a preferred oscillation phase (closest to the barrier) of the innermost ion.
What remains to be answered is why not every time separation equal to the period $T$ leads to transfer and why some time separations between the steps are two times the period $T$,
i.e. some steps are absent.

To provide an answer to this question it is essential to take into account also the dynamics of the other ions constituting the Coulomb crystals.
 
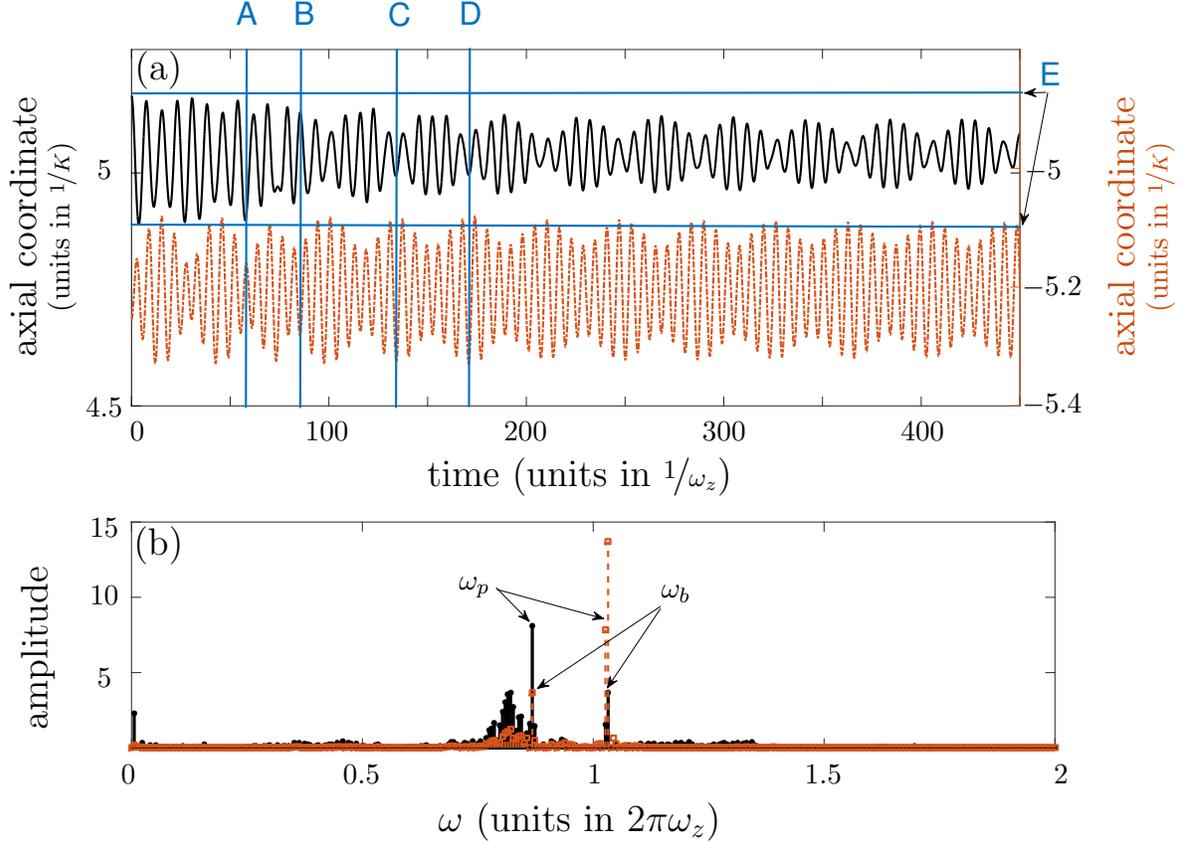
\begin{figure}[h!]
 \begin{tikzpicture}[node distance=5em, scale=0.4]
   \node at (0,0){
 \includegraphics[bb=0 0 1137 793,scale=0.4,keepaspectratio=true]{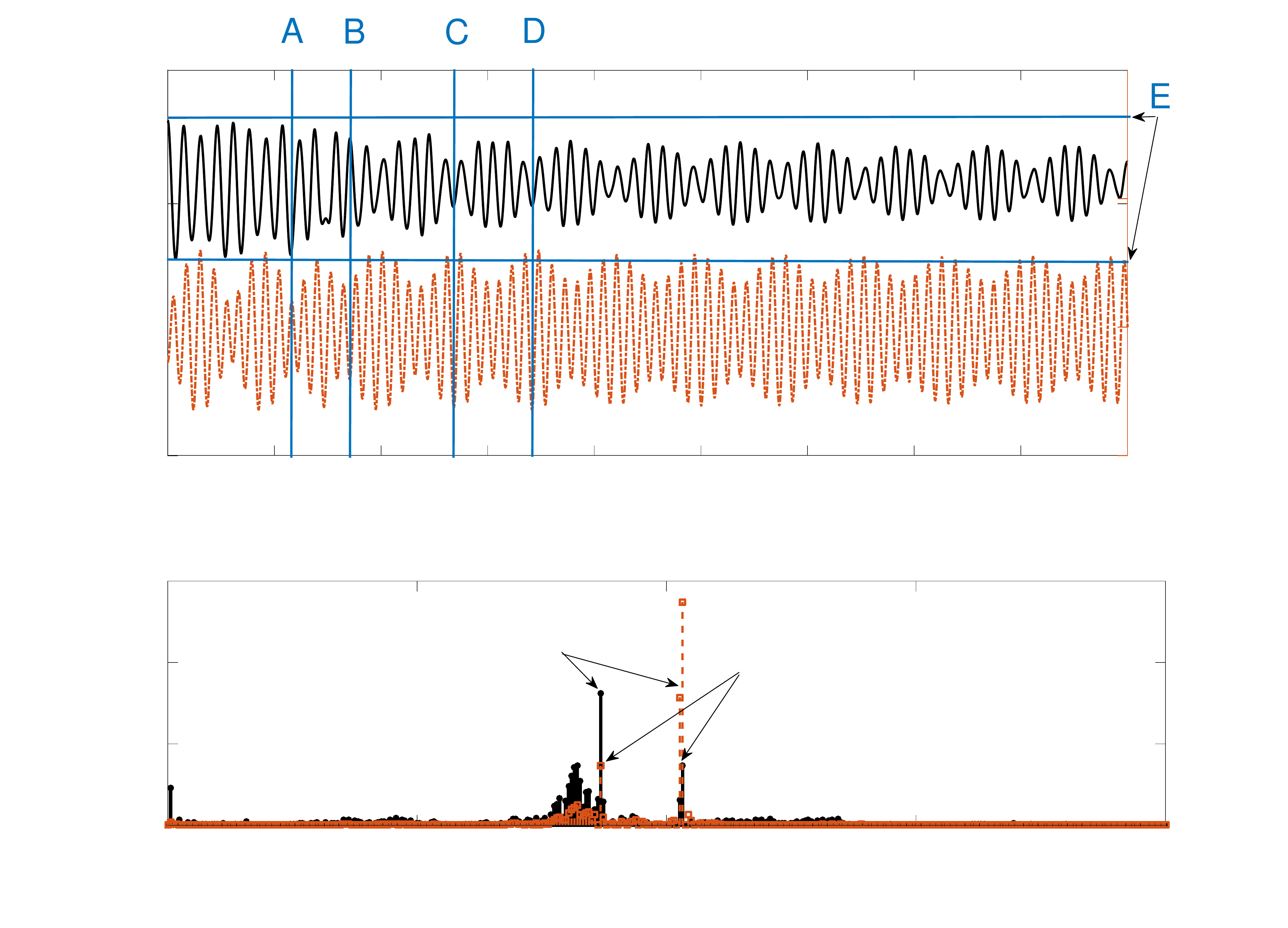}};
     \node at (-14,11.7) {\large (a)};
   \node at (-14,-4.1) {\large (b)};
 \node at(-14.7,-.3) {$0$};
 \node at(-8.4,-.3) {$100$};
\node at(-1.8,-.3) {$200$};
 \node at(4.8,-.3) {$300$};
\node at(11.4,-.3) {$400$};
 \node at(-15.8,.5){$4.5$};
 \node at(-15.8,8.4){$5$};
\node at(15.4,8.35){$-5$};
 \node at(15.7,4.5){$-5.2$};
 \node at(15.7,0.6){$-5.4$};
 \node at(-15.8,-8.5) {$5$};
 \node at(-15.8,-5.75) {$10$};
 \node at(-15.8,-3.5) {$15$};
 \node at(-15,-11.7) {$0$};
 \node at(-7.25,-11.7) {$0.5$};
 \node at(0.5,-11.7) {$1$};
 \node at(8.1,-11.7) {$1.5$};
\node at(16,-11.7) {$2$};
 \node at(-3.5,-5.5){$\omega_p$};
 \node at(3.2,-5.7) {$\omega_b$};
  \node at(0,-1.8) {\large time (units in $\nicefrac{1}{\omega_z}$)};
 \node[rotate=90] at(-18.5,6.5){\large axial coordinate};
\node[rotate=90] at(-17.2,6.5) {\small(units in $\nicefrac{1}{K}$)};
\node[rotate=90] at(18,6.5){\large {\color{redi}axial coordinate}};
\node[rotate=90] at(19.3,6.5) {\small{\color{redi}(units in $\nicefrac{1}{K}$)}};
 \node at(0,-13.3) {\large $\omega$ (units in $2\pi\omega_z$)};
 \node[rotate=90] at(-18,-7.5){\large amplitude};
 \end{tikzpicture}
\caption{(a) CM motion of the two crystals (large crystal, solid black line and small crystal, red dashed line) for the
zig-zag configuration with $C_f= 3.3403$. The vertical lines A,B,C,D mark the time instants during 
the dynamics corresponding to qualitatively different positions of the two Coulomb crystals. The horizontal lines E denote the initial amplitude of the
CM motion of the large crystal.
b) The corresponding Fourier spectra for the large (black- $\omega_{p}=0.868$ and $\omega_{b}=1.0264$) 
and the small crystal (red-- $\omega_{p}\approx1.0323$ and $\omega_{b}\approx0.868$).}
\label{Fig.:COM_big}
\end{figure}

 After the quench all ions constituting the two Coulomb crystals move towards the barrier, as it can be seen by inspecting their CM  motion (Fig.~\ref{Fig.:COM_big} (a)).
 Since the two crystals interact via repulsive Coulomb forces, the repulsion exerted by the small crystal hinders the transfer of the innermost ion of the large crystal.
 If the energy introduced by the quench is high enough to overcome the Coulomb interaction and the barrier, the innermost ion travels, otherwise the two crystals start
 to oscillate without any ion transfer, as depicted in their CM dynamics (Fig.~\ref{Fig.:COM_big} (a)). 
The Fourier analysis of this CM motion shows that the frequencies of the left and right crystals differ (Fig.~\ref{Fig.:COM_big} (b)) due to their different sizes. 
  Therefore there is a variety of  possibilities for the position of the two crystals during the time evolution (see Fig.~\ref{Fig.:COM_big}(a)):
(i) they can both be close to the barrier (line A in Fig.~\ref{Fig.:COM_big}(a)),(ii) both can be far away from the barrier (line B in Fig.~\ref{Fig.:COM_big}(a)),(iii)
the large crystal can be close to the barrier and the small far away (line C in Fig.~\ref{Fig.:COM_big}(a)) or (iv) the opposite of (iii) (line D in Fig.~\ref{Fig.:COM_big}(a)).
Obviously the case, with the small crystal being far and the large crystal being close to the barrier (case C) is optimal for the transfer, both in terms of energy
and spatial configuration.

  It turns out that the two CM frequencies depend also on the value of the barrier height, thus the time instant when the
 optimal conditions are fulfilled changes slightly with $C_f$, leading to different times for ion transfer appearing as different steps in Fig. \ref{Fig.:period}.
 In the same line of arguments, it is possible, depending on $C_f$ that the innermost ion is close to the barrier when the CM of the larger crystal is not and thus
 due to the lack of energy the ion transfer is prohibited, leading to some steps being skipped and the subsequent steps appearing after a time $2T$ (Fig. \ref{Fig.:period}).
  In Fig.~\ref{Fig.:COM_big} (b) we also see, that apart from the main frequencies $\omega_{p}$ for the two crystals, there also exist 
 additional beating frequencies $\omega_{b}$. This results in
 an amplitude modulation (see Fig.~\ref{Fig.:COM_big} (a)) which also influences the transfer dynamics.Interestingly enough the pairs of $\omega_p, \omega_b$ frequencies
 for the small and the large crystal are approximately degenerate, a fact that could be attributed to their Coulomb coupling.

Another feature of the CM dynamics is the damping of the oscillations of the big crystal with time
 (Fig.~\ref{Fig.:COM_big}(a) lines E) limiting the time available for ion transfer and giving rise to the observed gaps of Figs.~\ref{Fig:point_in_time},\ref{Fig.:begin_zigzag}. 
 The origins of this damping are the repulsive interactions between the two crystals, the transfer of energy in the radial directions
 (here especially in the less confining direction) and the mode coupling between the CM modes and other modes, due to the inherent nonlinearity of the system.
 
  In contrast to the above discussion, if the energy introduced by the quench is large enough, the details of the dynamics of the CM motion seize to rule the transfer process of an ion, 
 resulting in a smooth behaviour of the time instant for an ion to cross the barrier as a function of $C_f$ (Fig.~\ref{Fig:point_in_time}).

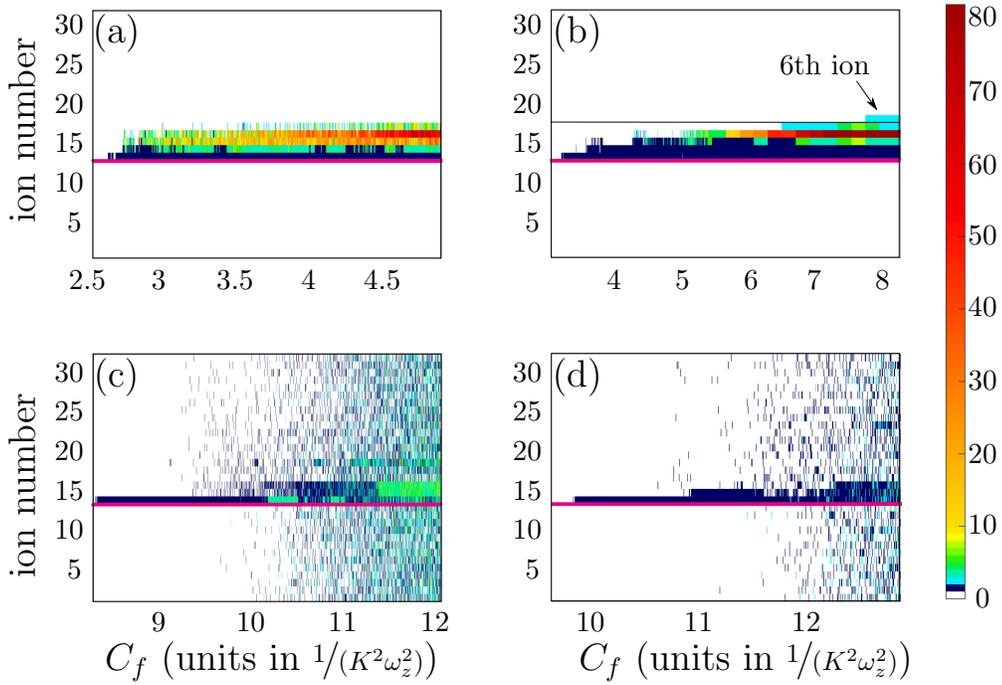
\begin{figure}[h!]
 \begin{tikzpicture}[node distance=5em, scale=0.35]
  \node at (0,0){
 \includegraphics[bb=0 0 1103 747,scale=0.35,keepaspectratio=true]{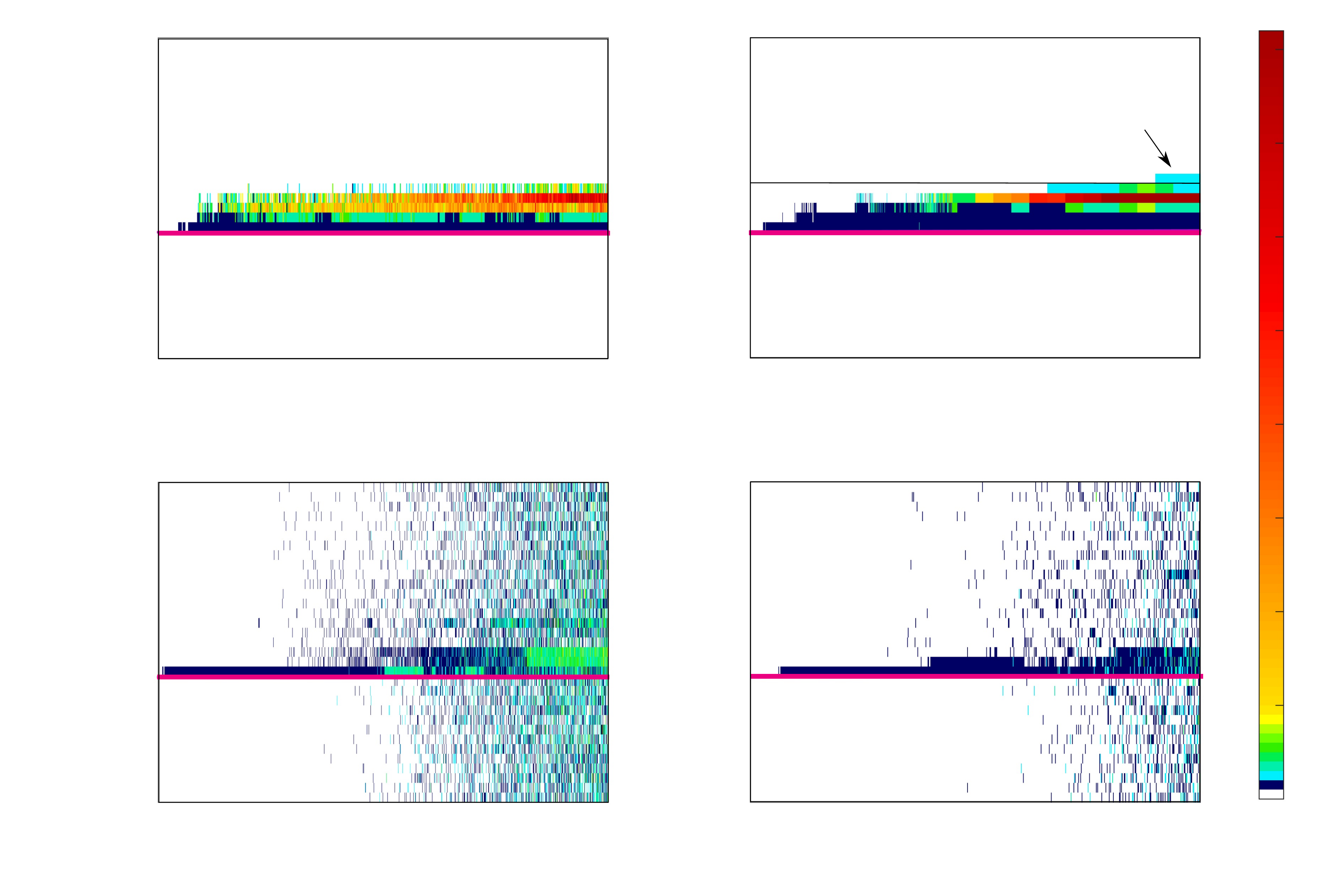}};
   
   \node at (-13.9,11.2) {\large (a)};
   \node at (3.6,11.2) {\large (b)};
   \node at (-13.9,-1.9) {\large (c)};
   \node at (3.6,-1.9) {\large (d)};
 \node at(-8.,-12.7) {\large $C_f$ (units in $\nicefrac{1}{(K^2\omega_z^2)}$)};
\node at(10,-12.7) {\large $C_f$ (units in $\nicefrac{1}{(K^2\omega_z^2)}$)};
 \node[rotate=90] at (-17.5,7.5){\large ion number}; 
 \node[rotate=90] at (-17.5,-6){\large ion number};
 \node at(-15,1.8){$2.5$};
 \node at(-12.3,1.8){$3$};
 \node at(-9.4,1.8){$3.5$};
 \node at(-6.6,1.8){$4$};
 \node at(-3.8,1.8){$4.5$};

 \node at(5,1.8){$4$};
 \node at(7.6,1.8){$5$};
 \node at(10.2,1.8){$6$};
 \node at(12.6,1.8){$7$};
 \node at(15.2,1.8){$8$};
 \node at(-12.3,-11.2){$9$};
 \node at(-8.8,-11.2){$10$};
 \node at(-5.3,-11.2){$11$};
 \node at(-1.8,-11.2){$12$};
 \node at(4.1,-11.2){$10$};
 \node at(8.2,-11.2){$11$};
 \node at(12.3,-11.2){$12$};
  \node at(-15.5,4){$5$};
 \node at(-15.7,5.5){$10$};
 \node at(-15.7,7){$15$};
 \node at(-15.7,8.5){$20$};
 \node at(-15.7,10){$25$};
 \node at(-15.7,11.5){$30$};
 \node at(13,10){$6$th ion};
  \node at(-15.5,-9.2){$5$};
 \node at(-15.7,-7.7){$10$};
 \node at(-15.7,-6.2){$15$};
 \node at(-15.7,-4.7){$20$};
 \node at(-15.7,-3.2){$25$};
 \node at(-15.7,-1.7){$30$};
 
   \node at(1.9,4){$5$};
 \node at(1.7,5.5){$10$};
 \node at(1.7,7){$15$};
 \node at(1.7,8.5){$20$};
 \node at(1.7,10){$25$};
 \node at(1.7,11.5){$30$};

  \node at(1.9,-9.2){$5$};
 \node at(1.7,-7.7){$10$};
 \node at(1.7,-6.2){$15$};
 \node at(1.7,-4.7){$20$};
 \node at(1.7,-3.2){$25$};
 \node at(1.7,-1.7){$30$};
 
 \node at(19.,-10.3){$0$};
 \node at(19,-7.5){$10$};
 \node at(19,-4.8){$20$};
 \node at(19,-2){$30$};
 \node at(19,.8){$40$};
 \node at(19,3.5){$50$};
 \node at(19,6.2){$60$};
 \node at(19,9){$70$};
 \node at(19,11.8){$80$};
\end{tikzpicture}
\caption{Number of transfers per ion as a function of  $C_f$ for a) linear chains b) zig-zag chains c) circles d) spheres. The pink line separates the ions of the small 
from the ions of the large crystal.}
\label{Fig.:Karte}
\end{figure}

So far we have focused on the first time instant at which an ion is transferred from one well to the other. But as already mentioned an ion can travel back and forth
thereby passing the potential barrier several times. To extract information on the number of transfers per ion, we have sorted the ions in their initial GS configuration according 
to their positions in the axial direction in increasing order and we have counted for each one the number of transfers occurring in the time interval considered for the dynamical evolution. 
Using this convention, the first 13 ions are at $t=0$ in the small and the next 20 in the big crystal.
In Fig.~\ref{Fig.:Karte} the results for the different initial configurations (Fig.~\ref{Fig.initial}) are shown.

We have already seen (Fig. \ref{Fig:point_in_time}) that for the linear and the zig-zag configurations, in the considered region of $C_f$, only five ions travel.
These are the ions which are located in the big crystal closest to 
the barrier (ion numbers 14 to 19). As we observe (Fig. \ref{Fig.:Karte} (a), (b)) 
these 5 ions travel  several times forth and back over the barrier during our simulation time,  whereas none of the other ions in the two crystals ever crosses the barrier.
The reason for this is the strict confinement in the radial direction for these two cases which  is especially true
for  the linear chains. For the zig-zag case, as long as the ion order in the axial direction is preserved the exclusive transfer of only 5 ions also holds,
but for larger values of $C_f>6.5$ (where the strict axial order is destroyed) further ions do also get the possibility of being transferred
(compare fig.~\ref{Fig.:Karte} (b)).

In contrast to the above, for the cases of the  circle and sphere configurations  (Fig. \ref{Fig.:Karte} (c), (d)) all ions can be transferred. The low aspect ratios  ($\alpha=1$) of the radial confining potentials 
of the circle and the spherical configurations enable  rotations of the whole crystals, as well as rearrangements with respect to the order of the ions constituting them.
Thus in the course of the dynamics,
the order of the ions in the axial direction changes and different ions are located at different times 
closer to barrier, resulting for larger $C_f$  in a nearly uniform distribution of the number of transfers among the ions of the  two crystals. 
 Having understood the basic features of the ion transfer processes, let us now examine how these do affect the order and the structure of the
two involved Coulomb crystals.

\subsection{Crystalline order}

The ion transfer processes discussed above yield a complex non-equilibrium dynamics of the two resulting Coulomb crystals 
involving reordering processes and the emergence of structural disorder. In order to characterize and analyze the order of the resulting crystals we make use of
a  measure based on the Voronoi diagrams introduced in \cite{Prep1985, mine}.
The time evolution of this measure has been proven to capture well the change in the crystalline order during the dynamics \cite{mine}.

The two Coulomb crystals have different sizes causing them to behave differently. For each crystal located in a certain well we determine its Voronoi measure value, taking also into
account that the number of ions per well (crystal) changes in time due to the ion transfer processes.
 Considering also the factor of dimensionality we arrive at the following definition  of our Voronoi measure $\Omega$

 \begin{equation}
 \Omega(t)=\gamma\frac{1}{N}\sum_i\Biggl(\frac{r_{ij}(t)}{2}\Biggr) ^d
\end{equation}
\label{Omega}
 where  $N$ is the number of ions in the well and $r_{ij}$ is the distance of each particle $i$ from its nearest neighbor $j$.
This measure corresponds to the sum of the areas of circles centered in each ion $i$, with a diameter equal to $r_{ij}$, i.e. the sum of
areas in which only one ion can be found. 
The variables $d$ and $\gamma$ depend on the dimensionality of the configurations.
For the linear chains (1D)  we have that $d=1$ and $\gamma=1$, for the zig-zag chains and the circular structures (2D) $d=2$ and $\gamma=\pi$  and for
the spherical configurations $d=3$ and $\gamma=4\pi/3$.

In the course of the dynamics the system of ions alternates between regular and irregular ion configurations resulting in an alternating Voronoi measure.
The regular structures which give visually the impression of order lead to distinct minima in the time evolution of the Voronoi measure,
whereas irregular structures lead to larger values of $ \Omega(t)$ \cite{mine}.

\begin{figure}[h!]
 \begin{tikzpicture}[node distance=5em, scale=0.15]
     \node at (0,0){
 \includegraphics[bb=0 0 2550 1402,scale=0.15,keepaspectratio=true]{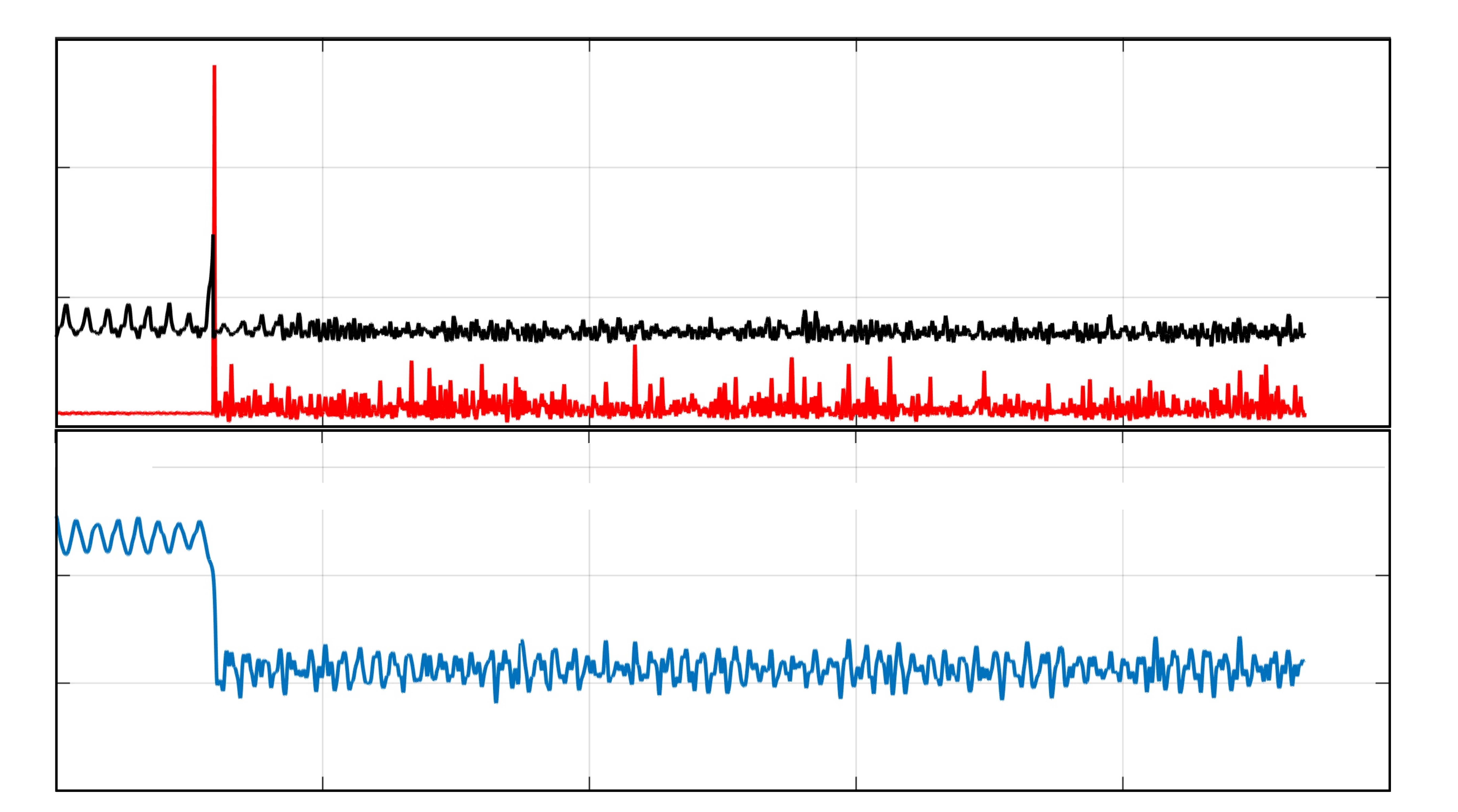}};
  
   \node at (-40.,12.7) {\large (a)};
   \node at (-40,-7.4) {\large (b)};
 \node[rotate=90] at(-52,6.) {\large $\Omega$};
 \node[rotate=90] at(-49,6) {(units in $\nicefrac{1}{K^d}$)};
 \node at(-8,-30){\large time (units in $\nicefrac{1}{\omega_z}$)};
\node[rotate=90] at(-52.6,-16){\large axial coordinate};
\node[rotate=90] at(-49.6,-16) {(units in $\nicefrac{1}{K}$)};
 \node at(-28.7,-26){$100$};
 \node at(-14.6,-26){$200$};
 \node at(-1,-26){$300$};
 \node at(12.5,-26){$400$};
 \node at(26.5,-26){$500$};
 \node at(-42,-26){$0$};
 \node at(-44.2,-18){$-2$};
 \node at(-43.4,-12.5){$0$};
 \node at(-43.4,-7){$2$};
 \node at(-45,2){$0.05$};
 \node at(-44,8.5){$0.1$};

\end{tikzpicture}
 \caption{ The time evolution of (a) the Voronoi measure ($\Omega$) of the two ion crystals (large ion crystal- black and small ion crystal- red) and (b) the axial position of the
 innermost ion for the zig-zag configuration and $C_f=3.34$}
 \label{fig:Voronoi}
\end{figure}
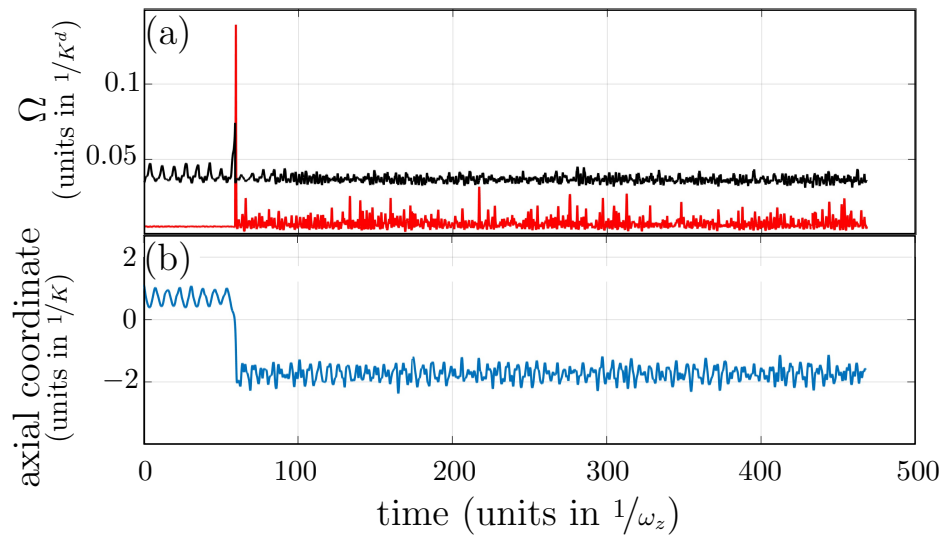

 An example of the time evolution of the Voronoi measure $\Omega(t)$ after the quench of the barrier height
is shown in Fig.~\ref{fig:Voronoi} (a)  for the zig-zag Coulomb crystals (Fig.~\ref{Fig.initial} (b))
and for the final quench value $C_f=3.34$ for which a single ion is transfered at $t\approx 59$ 
 (Figs.~\ref{Fig:point_in_time} (b), \ref{Fig.:begin_zigzag}). For comparison we also present the time evolution
of the axial coordinate of the traveling ion (Fig.~\ref{fig:Voronoi} (b)). We observe that initially (i.e. before the ion transfer)
the Voronoi measure exhibits regular oscillations with a much larger amplitude for the large crystal compared to the  smaller one.
This can be attributed to the larger shell diameter of the former which allows for larger deformations (i.e. compressions and expansions).

At the time instant when the ion crosses the barrier (Fig.~\ref{fig:Voronoi} (b)) the Voronoi values of 
both the large and the  small crystal exhibit a prominent peak  (Fig.~\ref{fig:Voronoi} (a)) resulting from the change in the 
number of ions per crystal and the fact that the distance of the traveling ion from its nearest neighbors maximizes
when it crosses the barrier. After the ion transfer the Voronoi measure of the small crystal performs highly irregular oscillations with an 
increased  amplitude, pointing to the irregular and disordered dynamics of the ions constituting the crystal (crystal melting).
In contrast, the oscillation amplitude of the Voronoi measure for the large crystal decreases after the ion transfer due to the increment of available space
for the ion dynamics and the substantial loss of energy caused by the loss of the highly energetic traveling ion. Similarly to the case of the 
smaller crystal the oscillations after the transfer become more irregular involving multiple frequencies.

These results suggest that the Voronoi measure $\Omega(t)$ and especially its oscillation amplitude encapsulates 
substantial information on the out-of-equilibrium many-body ion dynamics following the quench of the barrier height. Nevertheless, 
in order to proceed to an analysis of the  crystalline order as a function of the final quench parameter $C_f$ we would like to have a single value
characterizing each time series. A measure related closely to the average oscillation amplitude of $\Omega(t)$ (i.e. capturing well the
average dynamics of the crystals) is its standard deviation $\Delta\Omega$ in time.

\begin{figure}[h!] \begin{tikzpicture}[node distance=5em, scale=0.35]
      \node at (0,0){
 \includegraphics[bb=0 0 1151 793,scale=0.35,keepaspectratio=true]{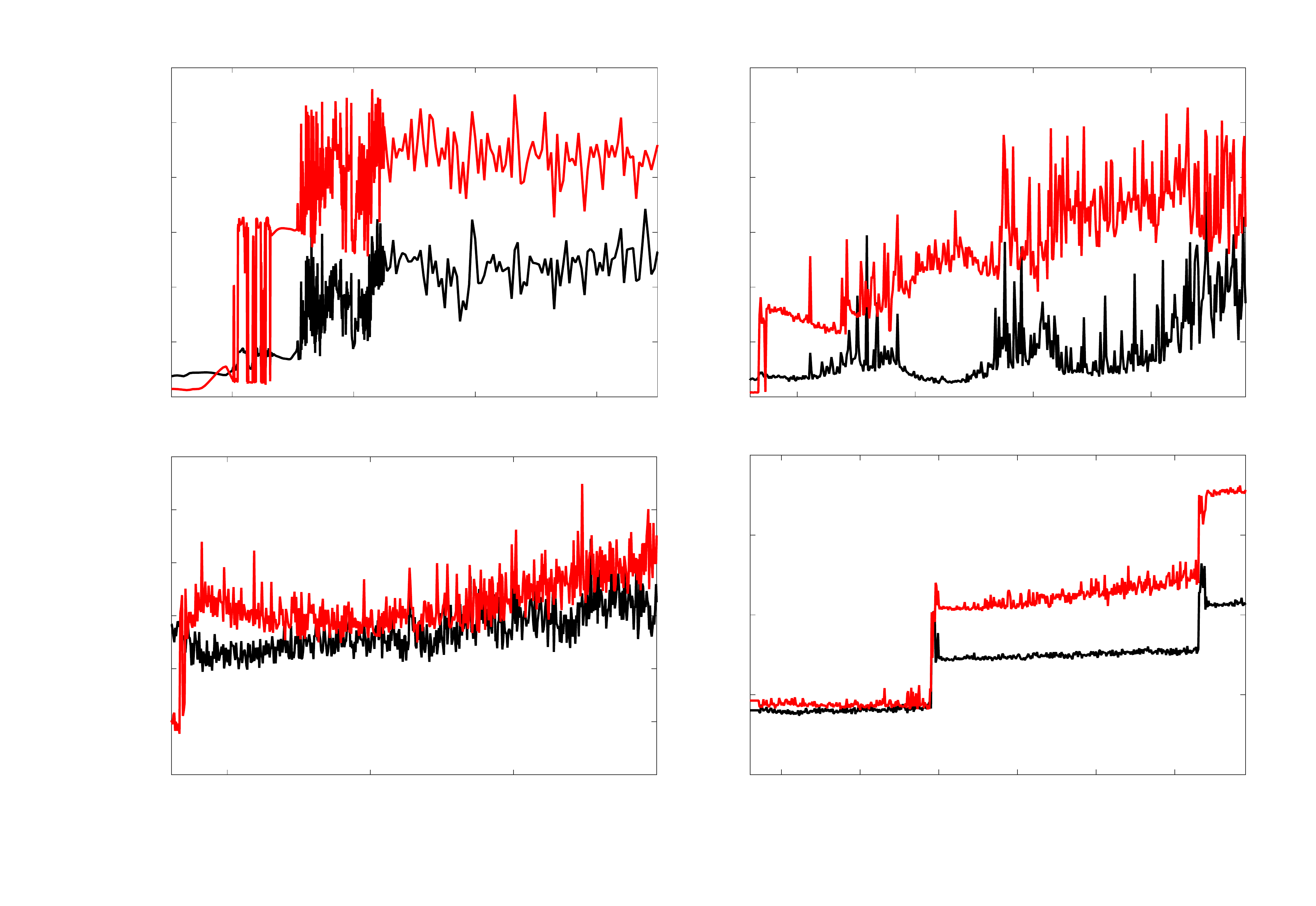}};

   \node at (-14.,10.7) {\large (a)};
   \node at (4,10.7) {\large (b)};
   \node at (-14.,-1.2) {\large (c)};
   \node at (4,-1.2) {\large (d)};
 \node[rotate=90] at(-19,7) {$\Delta\Omega$ (units in $\nicefrac{1}{K^d}$)};
  \node[rotate=90] at(-19,-5) {$\Delta\Omega$ (units in $\nicefrac{1}{K^d}$)};
\node at(-7.5,-13) {\large $C_f$ (units in $\nicefrac{1}{(K^2\omega_z^2)}$)};
\node at(10,-13) {\large $C_f$ (units in $\nicefrac{1}{(K^2\omega_z^2)}$)};

 \node at(-16.,1.5){$0$};
 \node at(-16.5,5){$0.02$};
 \node at(-16.5,8.5){$0.04$};
 \node at(-16.5,12){$0.06$};
 
\node at(-16.5,-6.8){$0.2$};
 \node at(-16.5,-3.3){$0.4$};
 \node at(-16.5,-.2){$0.6$};
 \node at(-16,-10.){$0$};
 \node at(2,1.5){$0$};
\node at(2,5.2){$2$};
  \node at(2,8.7){$4$};
 \node at(2,12){$6$};
 \node at(1.5,-10.){$0.5$};
 \node at(2,-7.5){$1$};
\node at(1.5,-5){$1.5$};
 \node at(2,-2.5){$2$};
 \node at(1.5,0){$2.5$};
 \node at(-12.7,1.){$2.6$};
\node at(-9.3,1){$2.8$};
 \node at(-5.5,1){$3$};
 \node at(-1.8,1){$3.2$};
 \node at(4.5,1){$3.5$};
 \node at(8,1){$4$};
 \node at(11.7,1){$4.5$};
 \node at(15.3,1){$5$};
 \node at(-13.3,-11){$8.5$};
 \node at(-8.8,-11){$9$};
 \node at(-4.5,-11){$9.5$};
 \node at(0,-11){$10$};
 \node at(3.8,-11){$10$};
 \node at(8.8,-11){$11$};
 \node at(18.3,-11){$13$};
 \node at(13.8,-11){$12$};
 \node at(4,12.5){\tiny$\times 10^{-3}$};
\end{tikzpicture}
 \caption{Standard deviation of $\Omega$ as a function of $C_f$ for (a) linear chain, (b) zig-zag, (c) circle and (d) sphere configurations (black line for the large and red line for the small crystal)}
\label{Fig.:Voronoi_all}
\end{figure}
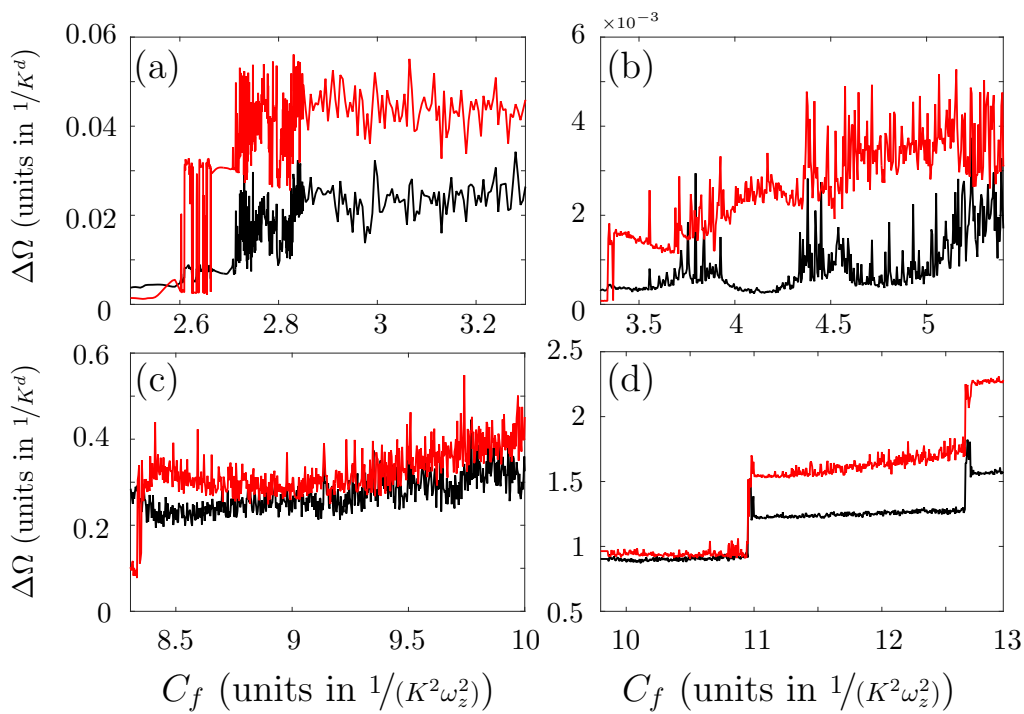

The resulting values for the standard deviations $\Delta\Omega$ as a function of $C_f$ are shown in Fig.
~\ref{Fig.:Voronoi_all} for the different trapping potentials examined (Fig.~\ref{Fig.initial}).
The standard deviation $\Delta\Omega$ for the one (Fig.~\ref{Fig.:Voronoi_all} (a)) and two-dimensional (Figs.~\ref{Fig.:Voronoi_all} (b),(c)) 
ion configurations are highly irregular but for 
the three-dimensional configuration (Fig.~\ref{Fig.:Voronoi_all} (d)) it is rather well structured, giving immediate access to relevant information. 

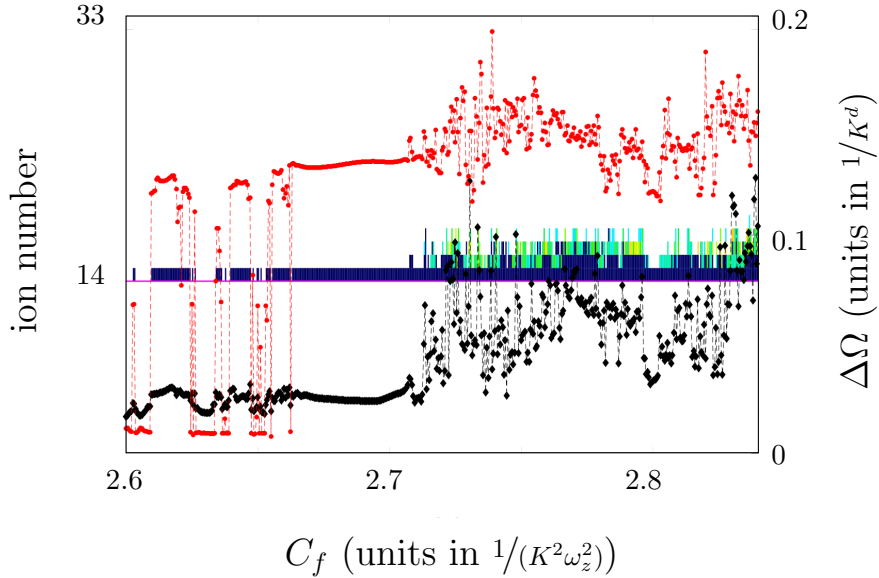
\begin{figure}[h!]
\vspace{-5cm}
 \begin{tikzpicture}[node distance=5em, scale=0.3]
  \node at (0,0){
 \includegraphics[bb=0 0 1516 1057,scale=0.3,keepaspectratio=true]{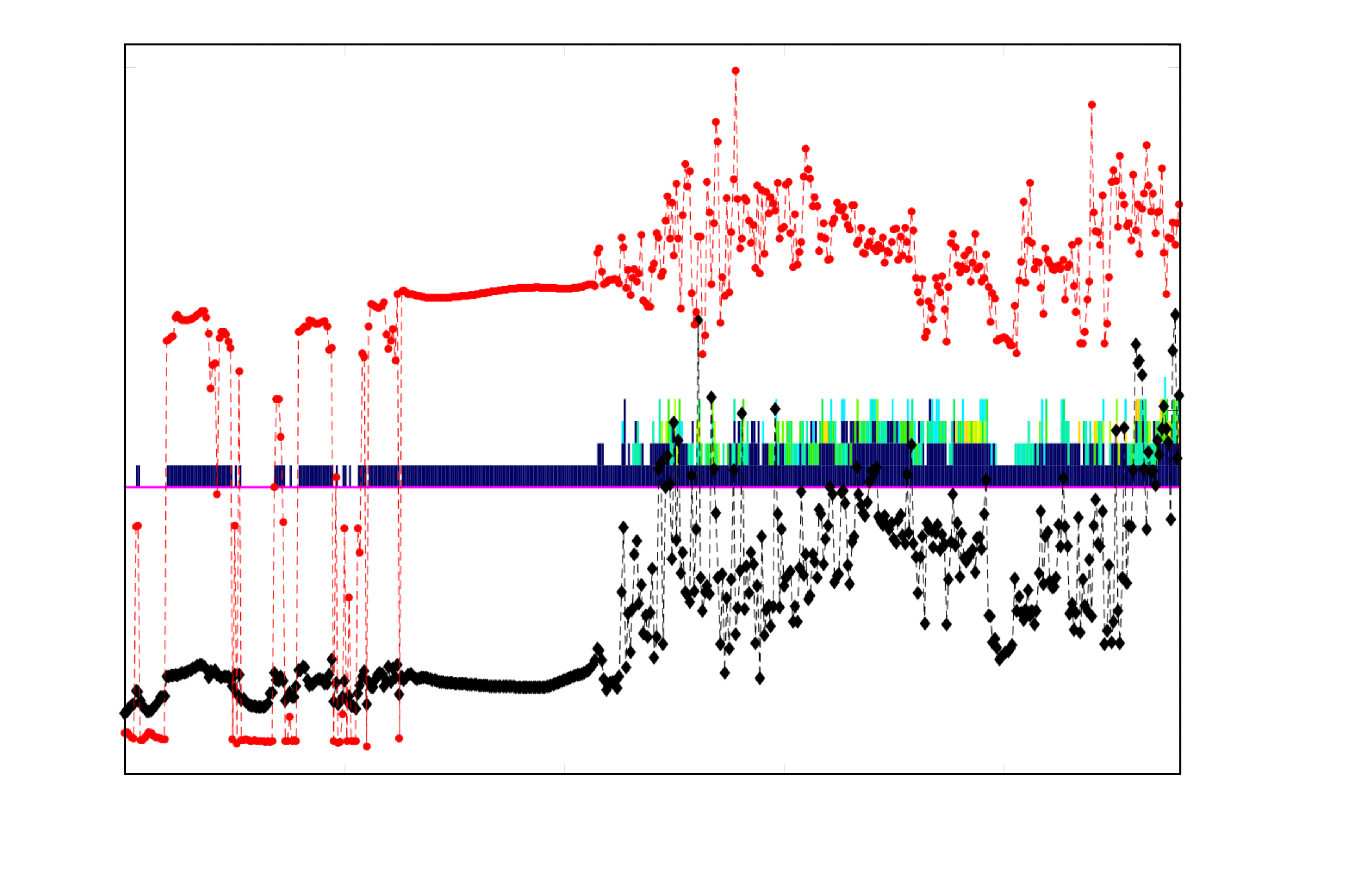}};
 \node[rotate=90] at(9,-6){\large $\Delta\Omega$ (units in $\nicefrac{1}{K^d}$)};
 \node[rotate=90] at(-28,-6) {\large ion number};
 \node at (-9,-20){\large $C_f$ (units in $\nicefrac{1}{(K^2\omega_z^2)}$)};
\node at(-25,-7.5){$14$};
 \node at(-25,4){$33$};
 \node at(-23.5,-16.6){$2.6$};
 \node at(-12.,-16.6){$2.7$};
 \node at(-0.5,-16.6){$2.8$};
 \node at(5.5,-15.5){$0$};
 \node at(6,-6){$0.1$};
 \node at(6,3.8){$0.2$};

\end{tikzpicture} 
\caption{Standard deviation and number of transfers per ion for the linear chain configuration in the range $C_f\in[2.6, 2.85]$ (black line for the large crystal and red line for the small crystal).} 
\label{fig:Chain_Karte_Voronoi}
\end{figure}

Focusing  on the noisy character encountered for example in the case of the linear chain, 
it turns out that many ion transfer processes occur.
 This can be inferred by an inspection of Fig.~\ref{fig:Chain_Karte_Voronoi}, where the  behaviour of $\Delta\Omega$
is compared to that of the number of transfers per ion as a function of $C_f$ in the interval $C_f\in[2.6, 2.85]$.
Clearly every small change in the transfer dynamics results in a substantial change in the standard deviation of the Voronoi measure 
$\Delta\Omega$, yielding the highly irregular pattern of the latter.

\begin{figure}[h!]
\vspace{-10cm}
\hspace{-1cm}
\begin{tikzpicture}[node distance=5em, scale=0.45]
   \node at (0,0){
 \includegraphics[bb=0 0 1516 1057,scale=0.45,keepaspectratio=true]{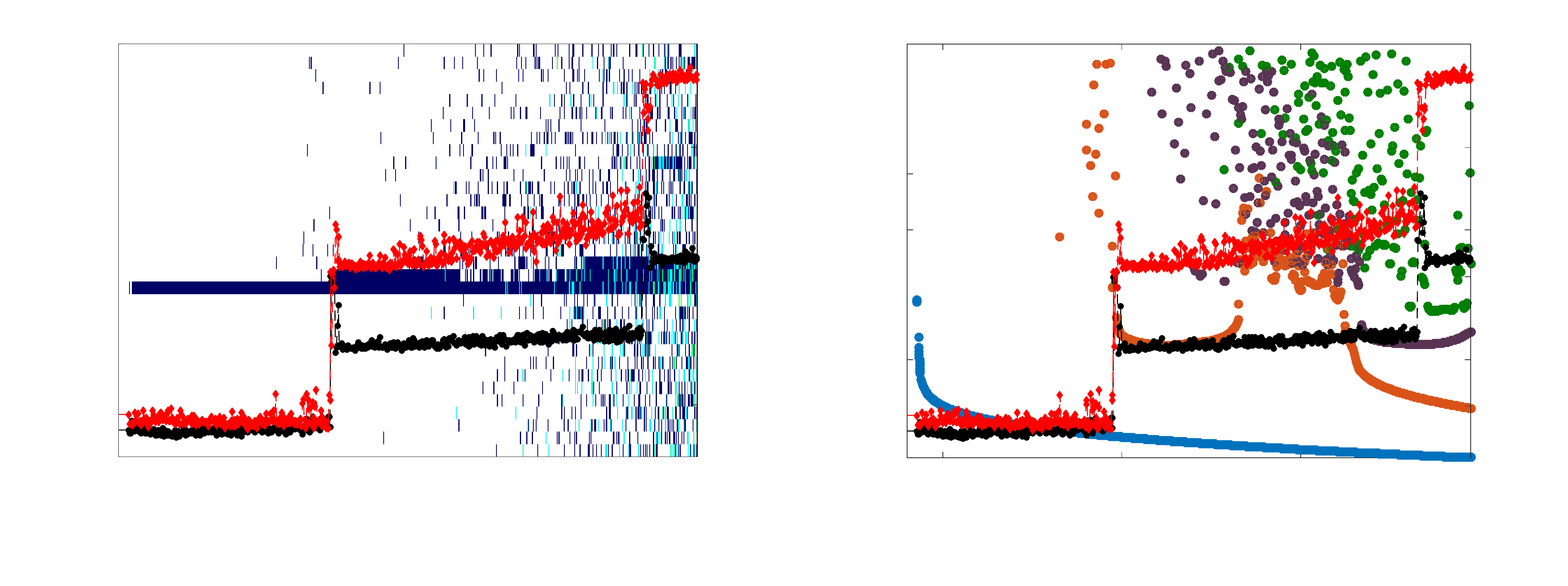}};
 \node[rotate=90] at(12.2,-10){\large$\Delta\Omega$ (units in $\nicefrac{1}{K^d}$)}; 
\node[rotate=90] at(-7.7,-10){\large$\Delta\Omega$ (units in $\nicefrac{1}{K^d}$)};
 \node[rotate=90] at(-26,-10) {\large ion number};
 \node at (-16,-17){\large $C_f$ (units in $\nicefrac{1}{(K^2\omega_z^2)}$)};
 \node[rotate=90] at(-5.1,-10) {\large time (units in $\nicefrac{1}{\omega_z}$)};
 \node at(4,-17) {\large $C_f$ (units in $\nicefrac{1}{(K^2\omega_z^2)}$)};
 \node at(-24.3,-11){$14$};
 \node at(-24.3,-4.7){$33$};
 \node at(-22.8,-16.){$10$};
\node at(-13,-16){$11$};
 \node at(-3.3,-16){$10$};
\node at(6.3,-16){$11$};
 \node at(-8.6,-14){$1$};
 \node at(-8.6,-7.5){$2$};
 \node at(-4.1,-12.8){$10$};
 \node at(-4.1,-8.2){$50$};
\node at(-4.3,-4.7){$500$};
 \node at(11.3,-14){$1$};
\node at(11.3,-7.5){$2$};

\end{tikzpicture}\caption{Standard deviation of the spherical configuration (black line for the large and red line for the small crystal) combined with a) the number of transfers per ion and b) the time instant of the first transfer of ions (the first 4 traveling ions-dots)}
 \label{fig:Voronoi_Kugel}
\end{figure}
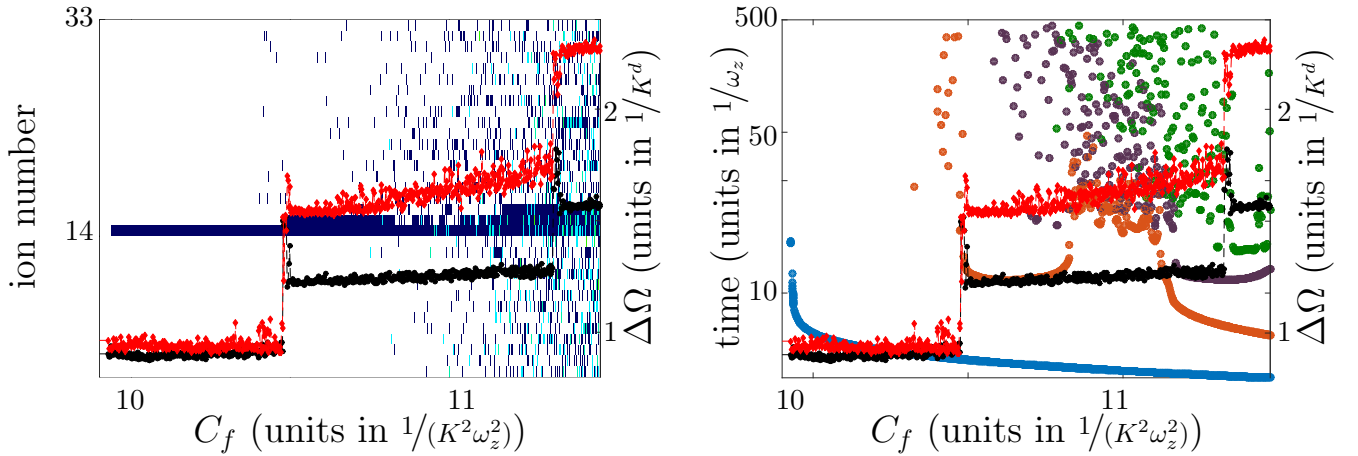
 For the  zig-zag and the circle configurations (Figs.~\ref{Fig.:Voronoi_all} (b), (c)) although there are intervals
of $\Delta\Omega$ exhibiting a smooth behaviour as a function of $C_f$ (especially for lower values of $C_f$) the overall pattern
is quite noisy as well resembling the case of the linear chain (Figs. ~\ref{Fig.:Voronoi_all} (a),~\ref{fig:Chain_Karte_Voronoi}).
In direct contrast, in the case of the  spherical configuration  (Fig.~\ref{Fig.:Voronoi_all} (d)) 
we  observe a rather regular behaviour of $\Delta\Omega$ as a function $C_f$,
allowing for extracting more directly information regarding the order of the Coulomb crystals involved. 

 Note here that among others the Voronoi measure and its standard deviation depend also on the dimensionality of the configurations (eq.~6) and the space available for motion.
Therefore, as it is obvious (Fig.~\ref{Fig.:Voronoi_all} (d)) the values of $\Delta\Omega$ for the three-dimensional configuration (spheres), where the available volume for the corresponding motion is much more enhanced, are
 orders of magnitude larger than that for the cases of lower dimensionality (Fig.~\ref{Fig.:Voronoi_all} (a)-(c)).
This results in the former being less sensitive to small deviations  yielding the overall regular pattern of $\Delta\Omega$ for the case of spheres.

 In particular we observe that $\Delta\Omega$ exhibits a quite smooth step-like behaviour interrupted by pronounced peaks
as $C_f$ increases.
In order to understand this behaviour we compare it to the $C_f$ dependence of the two quantities characterizing the ion transfer: the number of times each ion in the
Coulomb crystal travels back and forth between the two potential wells 
 (Fig.~\ref{fig:Voronoi_Kugel} (a))  and the time instant at which an arbitrary ion passes above the barrier for the first time (Fig.~\ref{fig:Voronoi_Kugel} (b)).
We observe that at most $C_f$ values for which an additional ion transfer occurs the
standard deviation of the Voronoi measure $\Delta\Omega$ possesses a peak, followed thereafter by the step-like behaviour of the other quantities
(Figs.~\ref{fig:Voronoi_Kugel} (a) and (b)). This can be interpreted as an increment of the structural disorder in the two crystals
induced by the increasing amount of ion transfer  processes and maximized each time a new ion gets transferred.
\subsection{Possible experimental realization}
Let us finally address the experimental realization of our setup, employing state of the art ion technology.
Typical experimental parameters for segmented Paul traps are $\omega_{rf}/2\pi=4.2-50 \text{MHz}$ and $U_{rf}=8-350\text{V}$ with applied
DC voltages in the axial direction  up to $10\text{V}$ \cite{Kauf2014,Hens2006}.
Depending on the ion species and trap design these result in  a radial confinement frequency $\omega/2\pi=1-5\text{MHz}$ and
in an axial confinement frequency $\omega_z/2\pi=0-5\text{MHz}$.
For the ion dynamics only the frequency ratio $\alpha=\frac{\omega}{\omega_z}$ matters and given the aforementioned frequency 
ranges the scenario studied in this work of $\alpha=8.25$ could be in principle realized ( choosing e.g $\omega/2\pi=4.5\text{MHz}$
and $\omega_z/2\pi=0.545\text{MHz}$). 
The parameters $z_0$ and $C$, determining the well positions and the barrier height respectively, depend on the axial DC voltage and the trap geometry,
thus realistic values for the former are of the order of 
$\unit{30}{\micro \meter} $ and for the latter up to $\unit{300}{\micro \meter}^2 \cdot \text{MHz}^2$.
Regarding the imaging of the ion configurations during their non-equilibrium dynamics, this could be achieved  by the use of fluorescence light detected by CCD cameras \cite{Miel2012,Kauf2014}.
\section{Conclusions}
\label{sec:outlook}
 We have explored the non-equilibrium dynamics of two Coulomb crystals of different sizes occupying 
the individual wells of a double-well potential, following a quench of the potential barrier height. 
The resulting complex dynamics is governed by ion transfer processes from one well to the other, depending on the quench amplitude.

The time instant at which an arbitrary ion passes the barrier for the first time shows an interesting step-like behaviour as a 
function of the quench amplitude. By analyzing the crystal dynamics we were able to explain the main features of this dependence. It turns out that the most crucial 
quantities determining whether ion transfer finally occurs are the center of mass motions of  both crystals and the oscillation frequencies  of the innermost ion. 

Following the ion transfer the dynamics of the two
crystals becomes rather irregular and characterized by structural disorder as well as reordering of the particles.
A good quantity to characterize the crystalline order is the so-called Voronoi measure. Its standard deviation in time reflects well the degree
of the structural disorder resulting from the quench and serves as a good indicator for the ion transfer processes.

A future work could investigate the case of two Coulomb crystals (with or without kinks), separated 
by a potential barrier, and aim at achieving a transport of medium-sized crystals (i.e. 10-100 ions).
Another promising direction for future research is the non-equilibrium dynamics of Coulomb crystals 
in multiple-well potentials resembling the optical lattices used in  studies of ultracold atoms.

\section*{Acknowledgments}
We thank Benno Liebchen for useful discussions and suggestions. A.K. thanks Stephan Klumpp for scientific discussions and support. 

\bibliography{AKtrappedIons}

\end{document}